\documentclass[aps,prb,preprint,showpacs,amsmath,amssymb,eqnarray,array]{revtex4}
\usepackage{anysize}
\marginsize{1.75cm}{1.75cm}{1.55cm}{1.55cm}
\usepackage{graphicx}
\usepackage{dcolumn}
\usepackage{bm}
\begin{document}

\title{Single-particle and collective excitations in quantum wires made up of vertically stacked quantum dots:
Zero magnetic field\\}
\author{Manvir S. Kushwaha}

\address
{\centerline {Department of Physics and Astronomy, Rice University, P.O. Box 1892, Houston, TX 77251, USA}}

\date{\today}

\begin{abstract}

We report on the theoretical investigation of the elementary electronic excitations in a quantum wires made up
of vertically stacked self-assembled InAs/GaAs quantum dots. The length scales (of a few nanometers) involved
in the experimental setups prompt us to consider an infinitely periodic system of two-dimensionally confined
(InAs) quantum dot layers separated by GaAs spacers. The resultant quantum wire is characterized by a
two-dimensional harmonic confining potential in the x-y plane and a periodic (Kronig-Penney) potential along
the z (or the growth) direction within the tight-binding approximation. Since the wells and barriers are formed
from two different materials, we employ the Bastard's boundary conditions in order to determine the
eigenfunctions along the z direction. These wave functions are then used to generate the Wannier functions,
which, in turn, constitute the legitimate Bloch functions that govern the electron dynamics along the direction
of periodicity. Thus the Bloch functions and the Hermite functions together characterize the whole system. We
then make use of the Bohm-Pines' (full) random-phase approximation in order to derive a general nonlocal,
dynamic dielectric function. Thus developed theoretical framework is then specified to work within a (lowest
miniband and) two-subband model that enables us to scrutinize the single-particle as well as collective
responses of the system. We compute and discuss the behavior of the eigenfunctions, band-widths, density of
states, Fermi energy, single-particle and collective excitations, and finally size up the importance of studying
the inverse dielectric function in relation with the quantum transport phenomena. It is remarkable to notice how
the variation in the barrier- and well-widths can allow us to tailor the excitation spectrum in the desired
energy range. Given the advantage of the vertically stacked quantum dots over the planar ones and the foreseen
applications in the single-electron devices and in the quantum computation, it is quite interesting and important
to explore the electronic, optical, and transport phenomena in such systems.

\end{abstract}
\pacs{73.21.-b, 73.22.-f, 73.63.-b, 78.67.-n}
\maketitle


\section{Introduction}

Even though the foundation of lower-dimensional electron systems was laid much before (than thought), the discovery
of quantum Hall effects (both integral and fractional) is known to have spurred the tremendous efforts to see the
consequent changes with the reduction in the system's dimensions from three to two, two to one, and one to zero.
These quasi-$n$-dimensional semiconductor heterostructures -- with $n$ ($=2$, 1, or 0) being the degree of freedom
-- are also known as quantum wells, quantum wires, and quantum dots in which the charge carriers exposed to external
probes such as electric and/or magnetic fields can exhibit unprecedented quantal effects that strongly modify their
behavior. Thanks to the advancements in the nanofabrication technology and electron lithography, these man-made
quantum structures have paved the way to much of the exotic (fundamental and applied) physics emerged during the
past two decades [1].

Years before vertically stacked self-assembled quantum dot molecules were synthesized, Sakaki had made a very
precise diagnosis of the relevant parameters defining the quantum transport in coupled quantum dot arrays in the
quantum limit [2]. This led him to propose the designing of such heterostructures with the optical phonon
scattering practically eliminated. It was argued that the optical phonon scattering could be completely ruled out
provided that the miniband (minigap) width is small (large) enough as compared to the optical phonon energy. It is
generally recognized that the absence of phonon scattering leads to dramatic effects in the transport phenomena.
For instance, it causes the efficient acceleration of electrons, which clearly improves the chances of the Bloch oscillations, so severely limited in the conventional systems.

The first vertically stacked self-assembled quantum dots were observed in InAs islands separated by GaAs spacer
layers along the growth direction in 1995 [3]. What followed is a long list of experimental [4-23] and theoretical
[24-40] works dealing with the elastic, electrical, electronic, and optical phenomena which led the researchers to
visualize a variety of solid-state devices [5]. It is already well-established that the strain due to the lattice
mismatch at the interfaces between two (different) semiconductors is the driving force for the growth of the
self-assembled quantum dots and is becoming known to play a crucial role in determining diverse electronic and
optical properties of the final system of vertically stacked quantum dots (VSQD).

The feasibility of varying the coupling strength (of the molecular bond) at will not only leads to some significant
applications in spintronics [15], quantum computation [21], solar cells [36], and  quantum optics [39] but also
creates a fertile ground for exploring new fundamental physics such as manipulating the spins [15] and electrical
control of electron g factors [37]. The most important and interesting (geometrical) aspect associated with the
vertically stacked quantum dot system is that it has provided, for the first time, a reasonable venue for
{\em reversing the trend} that began with the quest of diminishing dimensions in the mid 1980s. In other words, a
VSQD structure with a strong coupling along the growth direction offers a quasi-one-dimensional (Q1D) system made
up of quasi-zero-dimensional systems and hence the term ``reversing the trend".

Literature is a live witness that the plasmon, in classical as well as quantum systems, has always drawn more
attention than any other quasi-particle due, in fact, to its fundamental importance to the understanding of the
electronic, optical, and transport phenomena in condensed matter physics [1]. And yet, it is really surprising
that except for Ref. 34 no experimental and/or theoretical work on VSQD has, to the best of our knowledge, touched
upon the subject. The present paper helps fill that void. Here, we embark on a systematic route to the investigation
of the single-particle and collective excitations in the quantum wire made up of VSQD within a two-subband model in
the framework of Bohm-Pines' random-phase approximation (RPA) [41] in the absence of an applied magnetic field.

It is noteworthy that the theoretical development in quantum wires has long suffered from an intense controversy
over whether the system is best describable as Tomonaga-Luttinger liquid or as Fermi liquid. This issue was
elegantly resolved by Das Sarma and coworkers [42] who rigorously justified the use of Fermi-liquid-like theories
(such as the RPA) for describing the realistic quantum wires. It should be pointed out that our theoretical
framework differs crucially from Ref. 34 where the authors followed the effective-mass approximation [see Sec. II
for details]. For the extensive review on the plasmons in quantum wells, wires, dots, and their periodic counterparts,
the  reader is referred to Ref. 1.

The rest of the article is organized as follows. In Sec. II, we present the theoretical framework leading to the
derivation of nonlocal, dynamic, dielectric function, which is further diagnosed analytically to fully address
the solution of the problem and the related relevant aspects. In Sec. III, we discuss several illustrative examples
of, for example, excitation spectrum comprising of single-particle and collective excitations, the influence of the
variation of well- and barrier-thicknesses, and highlight the importance of studying the inverse dielectric function
in relation with the transport phenomena [43] in such quantum systems. Finally, we conclude our finding with specific
remarks regarding the interesting features worth adding to the problem in Sec. IV.

\section{Theoretical Framework}

\subsection{Eigenfunctions and eigenenergies}

We consider a periodic system of quasi-two-dimensional InAs islands of thickness $a$ separated by GaAs spacer layers
of thickness $b$. Each of the InAs island is constrained by a two-dimensional harmonic confining potential of the
form of $V(x)=\frac{1}{2}m^*\omega_0^2 (x^2+y^2)$ in the x-y plane. The z (i.e., the growth) direction is assumed
to be under the greater confinement potential, say, $V_c (z)$, that allows strong coupling between the InAs islands.
The small length scales and strong coupling cause the resultant structure (see Fig. 1) to mimic a realistic
quantum wire with a practically well-defined linear charge density ($n_{1D}$). Since the coupling strength and hence
the tunneling can be controlled by varying the thickness of the barrier between the quantum dot layers, we believe
that the use of the (traditional) tight-binding approximation (TBA) is quite justifiable [33]. Next, we consider the
resultant system with moderate tunneling (in the polarizability function) describable with the energy dispersion due
to tunneling being sinusoidal. Such a system as described above can be formally characterized by the eigenfunction

\begin{figure}[htbp]
\includegraphics*[width=8cm,height=9cm]{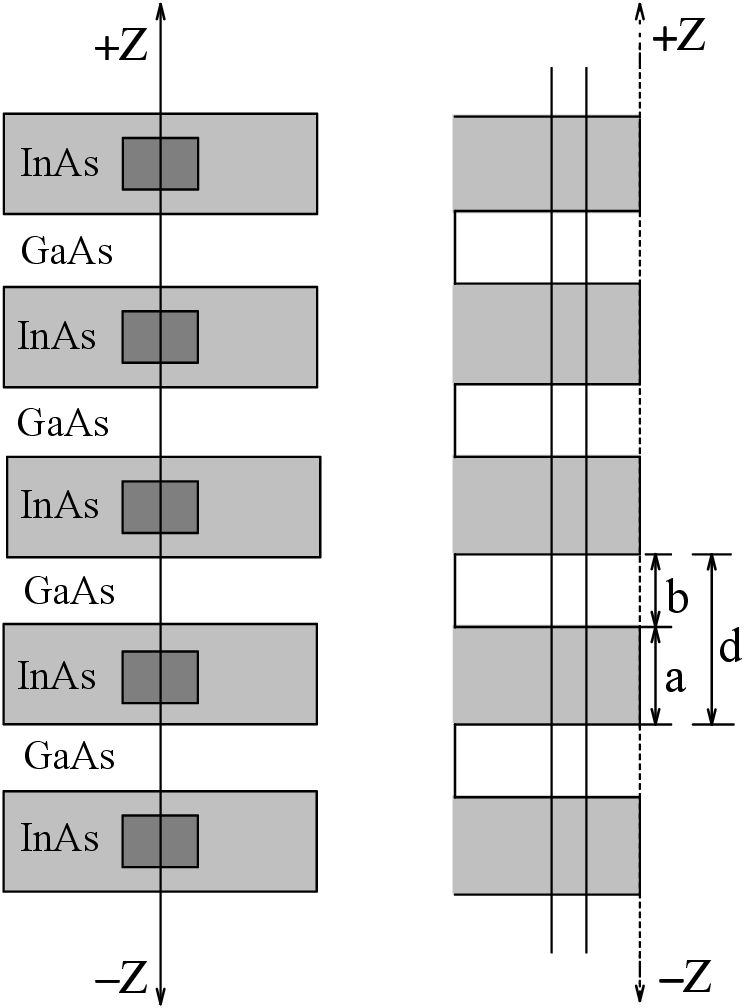}
\caption{Schematic of the quantum wire made up of an infinitely periodic system of InAs islands separated by GaAs
spacer layers (left panel). The right panel shows the Kronig-Penney periodic-potential simulation along the
growth direction. Here $a$ ($b$) is the well (barrier) width and $d=a+b$ is the period of the resultant system
making up quantum wire.}
\label{fig1}
\end{figure}

\begin{equation}
\psi (x,y,z)=\phi_{n_{x}}(x)\,\phi_{n_{x}}(y)\,\phi_t(z),
\end{equation}
where

\begin{equation}
\phi_{n_{s}}(s)= N_{n_{s}}\,e^{-s^2/2\ell^2_{c}}\, H_{n_{s}}(s/\ell_{c}),
\end{equation}
where $s\equiv x, y$, $\ell_{c}=\sqrt{\hbar/m^*\omega_{_0}}$ is the characteristic length, $\omega_{_0}$ is the
frequency of the harmonic oscillator, $N_n=1/\sqrt{\sqrt{\pi}\,2^n\,n!\,\ell_{c}}$ is the standard normalization
constant, and $H_{n_{s}}(s/\ell_{c})$ is the Hermite polynomial, and

\begin{equation}
\phi_{t}(z)= \frac{1}{\sqrt{N}}\,\sum_l\,e^{i\,k\,l\,d}\,\chi_{_{t}}(z-ld),
\end{equation}
where $d=a+b$ is the period, $k$ the Bloch vector, $\chi_{_{t}}(...)$ the well-known Wannier function, and the
eigenenergy

\begin{equation}
\epsilon_{knt}=(n+1)\,\hbar\omega_0 + \epsilon_t - \frac{W_t}{2}\, \cos(k\,d),
\end{equation}
where $t$ ($n=n_x+n_y$) is the miniband (subband) index, $\epsilon_t$ the energy of the $t-$th miniband and $W_t$
is the band-width defined as

\begin{equation}
W_t = - 4 \int^{+a/2}_{-a/2}\, dz\, \chi_{_{t}} (z)\, V_0\, \chi_{_{t}}(z-d),
\end{equation}
where we assume that the confining potential is a finite square well (Kronig-Penney potential) with a barrier
height $V_0$ and well-width $a$. Since $N$ (the number of quantum dot layers) is very large, the sum in Eq. (3)
can be written as integral according to the replacement rule: $\sum_k\, \rightarrow \,(N/L_{z})\, \int_{_{BZ}}\,
dk$. Note that Eq. (3) represents the tight-binding constraint which hypothesizes a little overlap between the
wave functions of different sites. There $\chi_{_{t}}(...)$, if normalized in the length of the lattice (or,
nearly enough, in infinite length), satisfies: $\int dz\,\chi^*_{_{t}}(z-nd)\,\chi_{_{t}}(z-ld)=\delta_{nl}$
and $\int dz\,\phi^*_{t}(z)\,\phi_{t}(z)=1$. Here $L_z=Nd$ is the total crystal length along the growth direction.

\subsection{Nonlocal, dynamic dielectric function}

We start with the general expression of the non-interacting single-particle density-density response function
(DDRF) $\chi^0 (...)$ given by [1]

\begin{equation}
\chi^{0} ({\bf r},{\bf r'};\omega)=\sum_{ij}\, \Lambda_{ij}\,\,
\psi^*_i ({\bf r})\,\psi_j ({\bf r})\,
\psi^*_j ({\bf r'})\,\psi_i ({\bf r'}),
\end{equation}
where ${\bf r}\equiv (x,y,z)$, the composite index $i,j\equiv k,n,t$, and $\Lambda_{ij}$ is defined as follows.

\begin{equation}
\Lambda_{ij}= 2\, \frac{f(\epsilon_i)-f(\epsilon_j)}{\epsilon_i-\epsilon_j+\hbar\omega^*},
\end{equation}
where $f(x)$ is the well-known Fermi distribution function. $\omega^*=\omega+i\gamma$ and small but nonzero $\gamma$ represents the adiabatic switching of the Coulomb interactions in the remote past. The factor of $2$ takes care of
the spin degeneracy.

Next, we write the induced particle density with the aid of the Kubo's correlation function [1]. The result is

\begin{eqnarray}
n_{in}(x, y, z;\omega)
& =& \int dx' \int dy' \int dz'\, \chi^0 (x, y, z; x', y', z';\omega)\,V_{tot}(x', y', z';\omega)\nonumber\\
& =& \int dx' \int dy' \int dz'\, \chi (x, y, z; x', y', z';\omega)\,V_{ex}(x', y', z';\omega),
\end{eqnarray}
where $V_{tot}=V_{ex}+V_{in}$ is the total potential, with $V_{ex}$ ($V_{in}$) as the external (induced) potential.
$\chi$ and $\chi^0$ are, respectively, the total and the single-particle DDRF and are related to each other through
an integral Dyson equation

\begin{equation}
\chi ({\bf r}, {\bf r'};\omega) = \chi^0({\bf r}, {\bf r'};\omega) + \int d{\bf r^{''}}\,\int d{\bf r^{'''}}\,
\chi^0({\bf r}, {\bf r^{''}};\omega)\, V_{ee}({\bf r^{''}}, {\bf r^{'''}})\,\chi({\bf r^{'''}}, {\bf r'};\omega),
\end{equation}
where $V_{ee}(...)$ represents the binary Coulomb interactions and is defined as

\begin{equation}
V_{ee}({\bf r}, {\bf r'})=\frac{e^2}{\epsilon_b}\,\frac{1}
{\mid (x-x')^2+(y-y')^2+(z-z')^2 \mid^{1/2}},
\end{equation}
where $-e$ ($e>0$) is the elementary charge and $\epsilon_b$ the background dielectric constant of the system.
Further, the induced potential in terms of the induced particle density is expressed as

\begin{equation}
V_{in}(x, y, z;\omega)=\int dx' \int dy' \int dz'\, V_{ee}(x-x', y-y', z-z')\,n_{in}(x', y', z';\omega)
\end{equation}
Equation (11), with the aid of Eqs. (1), (3), (6), (8), and (10), yields


\begin{eqnarray}
V_{in}(x, y, z;\omega)=
&& \frac{e^2}{\epsilon_b}\frac{1}{N^2}\,\sum_{k k'}\,\sum_{n n'}\,\sum_{t t'}\,\sum_{l l'}\,
   \Lambda_{\stackrel{nn'}{tt'}}(k,k';\omega)\,e^{iqld}\,e^{-iql'd}\nonumber\\
&& \times \, \int dx'\int dy'\int dz'\int dx''\int dy''\int dz'' \nonumber\\
&& \times \, \frac{1}{\mid (x-x')^2+(y-y')^2+(z-z')^2 \mid^{1/2}}\,\nonumber\\
&& \times \, \phi^*_{n_{x}}(x')\phi^*_{n_{y}}(y')\phi_{n'_{x}}(x')\phi_{n'_{y}}(y')
     \phi^*_{n'_{x}}(x'')\phi^*_{n'_{y}}(y'')\phi_{n_{x}}(x'')\phi_{n_{y}}(y'')\nonumber\\
&& \times \, \chi^*_{_{t}}(z'-ld)\chi_{_{t'}}(z'-ld)\chi^*_{_{t'}}(z''-l'd)\chi_{_{t}}(z''-l'd)\nonumber\\
&& \times \, V_{tot}(x'',y'',z'';\omega),
\end{eqnarray}
where $k'=k+q$ and $q$ is the momentum transfer. Next, we follow two steps: (i) multiply both sides of Eq. (12)
by $e^{-iq'z}$ and integrate with respect to z, and (ii) introduce, for convenience, a single-Fourier component
of the total potential to write $V_{tot}(x'', y'', z'';\omega)=e^{q'z''}\,V_{tot}(x'',y'',q';\omega)$. The result
is

\begin{eqnarray}
V_{in}(x, y, q';\omega)=
&& \frac{2\,e^2}{\epsilon_b\,N^2}\,\sum_{k'}\,\sum_{n n'}\,\sum_{t t'}\,\sum_{l l'}\,
   \Pi_{\stackrel{nn'}{tt'}}(k,k';\omega)\,e^{-i(q'-q)ld}\,e^{i(q'-q)l'd}\nonumber\\
&& \times \int dx'\int dy'\, K_0(q'\mid \overline{r}-\overline{r}'\mid)\,
   \phi^*_{n_{x}}(x')\phi^*_{n_{y}}(y')\phi_{n'_{x}}(x')\phi_{n'_{y}}(y')\nonumber\\
&& \times \int dx''\int dy''\, V_{tot}(x'',y''; q',\omega)\,
   \phi^*_{n'_{x}}(x'')\phi^*_{n'_{y}}(y'')\phi_{n_{x}}(x'')\phi_{n_{y}}(y'') \nonumber\\
&& \times \int dz'\, e^{-iq'(z'-ld)}\, \chi^*_{_{t}}(z'-ld)\chi_{_{t'}}(z'-ld) \nonumber\\
&& \times \int dz''\, e^{iq'(z''-l'd)}\,\chi^*_{_{t'}}(z''-l'd)\chi_{_{t}}(z''-l'd),
\end{eqnarray}
where $\overline{r}$ and $\overline{r}'$ are the 2D vectors in the x-y plane, $K_{_0} (x)$ is the zeroth-order
modified Bessel function of the second kind, which diverges as $-\ln (x)$ when $x \rightarrow 0$, and

\begin{equation}
\Pi_{\stackrel{nn'}{tt'}}(k,k';\omega)=
\sum_{k}\, \Lambda_{\stackrel{nn'}{tt'}}(k,k';\omega)=2\,
\sum_{k}\,\frac{f(\epsilon_{nt})-f(\epsilon_{n't'})}{\epsilon_{nt}-\epsilon_{n't'}+\hbar\omega^*}
\end{equation}
Now, it is convenient to simplify a few steps before we proceed further. (i) The last two integrals in Eq. (13)
can be written as: $\mid \int dz\, e^{-iq'z}\,\chi^*_{_{t}}(z)\chi_{_{t'}}(z) \mid^2$. (ii) The sum over $l'$ is
worked out as follows:

\begin{eqnarray}
\frac{1}{N}\,\sum_{l'}\,e^{i(q-q')(l'-l)d}
&=& \frac{1}{N\,d}\,\int d(l'd)\,e^{i(q-q')(l'-l)d}\nonumber\\
&=& \frac{1}{N\,d}\, 2\pi \,\delta (q-q')\,
= \frac{1}{L_z}\, 2\pi \,\delta (q-q')
\end{eqnarray}
(iii) The remaining sum over $l$ reads as follows: $\sum_l \, 1 = N$. This cancels the remaining $N$ in the
denominator. (iv) Then open up the sum over $k'$ to write, with the aid of Eq. (15),

\begin{eqnarray}
\sum_{k'} \,\Pi_{\stackrel{nn'}{tt'}}(k,k';\omega)\,\frac{2\pi}{L_z}\,\delta (q-q')
&&= \int dk'\, \Pi_{\stackrel{nn'}{tt'}}(k,k';\omega)\,\delta (k'-k-q')\nonumber\\
&&= \Pi_{\stackrel{nt}{n't'}}(k,k'=k+q';\omega)
\end{eqnarray}
(v) Finally, in the light of Eqs. $(14)-(16)$, one can safely replace $q'$ by $q$ on both sides of Eq. (13)
to write

\begin{eqnarray}
V_{in}(x, y, q;\omega)=
&& \frac{2\,e^2}{\epsilon_b}\,\sum_{n n'}\,\sum_{t t'}\,
   \Pi_{\stackrel{nn'}{tt'}}(k,k'=k+q;\omega)\nonumber\\
&& \times \int dx'\int dy'\, K_0(q\mid \overline{r}-\overline{r}'\mid)\,
   \phi^*_{n_{x}}(x')\phi^*_{n_{y}}(y')\phi_{n'_{x}}(x')\phi_{n'_{y}}(y')\nonumber\\
&& \times \int dx''\int dy''\, V_{tot}(x'',y''; q',\omega)\,
   \phi^*_{n'_{x}}(x'')\phi^*_{n'_{y}}(y'')\phi_{n_{x}}(x'')\phi_{n_{y}}(y'') \nonumber\\
&& \times \left |\int dz \, e^{-iqz}\, \chi^*_{_{t}}(z)\chi_{_{t'}}(z)\right |^2
\end{eqnarray}
Next, we take the matrix elements of both sides between the states $\mid m_x, m_y>\equiv \mid m>$ and
$\mid m'_x, m'_y>\equiv \mid m'>$ to write

\begin{eqnarray}
<m'\mid V_{in}(...)\mid m>\,
=\sum_{nn'}\,\sum_{tt'} && \Pi_{\stackrel{nn'}{tt'}}(k,k'=k+q;\omega)\, U_{nn'mm'}(q)\nonumber\\
&& \times \, S_{tt'}(q)\, <n'\mid V_{tot}(...)\mid n>,
\end{eqnarray}
where

\begin{equation}
U_{nn'mm'}(q)=\frac{2e^2}{\epsilon_b}\,\int d\overline{r}\int d\overline{r}'\,
\phi^*_{n}(\overline{r})\,\phi_{n'}(\overline{r})\,
K_0(q\mid \overline{r}-\overline{r}'\mid)\,
\phi^*_{m'}(\overline{r}')\,\phi_{m}(\overline{r}'),
\end{equation}
\begin{equation}
S_{tt'}(q)=\left |\int dz \, e^{-iqz}\, \chi^*_{_{t}}(z)\chi_{_{t'}}(z)\right |^2,
\end{equation}
and
\begin{equation}
<n'\mid V_{tot}(...)\mid n>\,=\int d\overline{r}\,
\phi^*_{n'}(\overline{r})\,V_{tot}(\overline{r}; q, \omega)\,\phi_{n}(\overline{r})
\end{equation}
Notice that the quantities in Eq. (18) have been defined in Eqs. $(19)-(21)$ in the condensed notations for the
sake of brevity. Here $U_{nn'mm'}(q)$ is the matrix element of the Fourier-transformed Coulombic interactions.
Next, let us invoke the condition of self-consistency [$V_{tot}=V_{ex}+V_{in}$] on Eq. (18) to write

\begin{eqnarray}
<m'\mid V_{ex}(...)\mid m>\,= \sum_{nn'}\, \left [\delta_{nm}\,\delta_{n'm'} -
\sum_{tt'}\,\Pi_{\stackrel{nn'}{tt'}}(k,k'=k+q;\omega)\,U_{nn'mm'}(q)\right .\nonumber\\
\left . \times \, S_{tt'}(q) \right ]\,<n'\mid V_{tot}(...)\mid n>
\end{eqnarray}
Here $\delta_{nm}$ is the Kronecker delta function. Now, since the external potential and the total potential
are correlated through the nonlocal, dynamic dielectric function $\epsilon (\overline{r}, \overline{r}';\omega)$
such as

\begin{equation}
V_{ex}(\overline{r}, \omega)=\int d\overline{r}'\, \epsilon (\overline{r}, \overline{r}';\omega)\,
V_{tot}(\overline{r}', \omega),
\end{equation}
we can easily deduce from Eq. (22) that the generalized nonlocal, dynamic dielectric function is given by

\begin{equation}
\epsilon_{nn'mm'} (q; \omega)=\delta_{nm}\,\delta_{n'm'} - \sum_{tt'}\,
\Pi_{\stackrel{nn'}{tt'}}(k,k'=k+q;\omega)\, S_{tt'}(q)\, U_{nn'mm'}(q)
\end{equation}
We call attention to the fact that Eq. (24) is the main result that could be exploited to compute a host of
electronic and optical properties of the system at hand. It also plays an important role in studying, for
example, the inelastic electron and light (or Raman) scattering and the transport properties if we choose
to work in terms of the inverse dielectric function $\epsilon^{-1}_{nn'mm'}(q; \omega)$ [1, 43]. As regards
the computation of the excitation spectrum comprised of the single-particle and collective excitations, the
zeros of the dielectric function (DF) and the poles of the inverse dielectric function (IDF) must yield
exactly identical results.

\subsection{Symmetry and degeneracy}

The problem of the 2D harmonic oscillator offers an opportunity to show the critical connection between the
symmetry and the degeneracy. To this end, we need to recognize that the subband index for the energy level
is $n=n_x+n_y$, where $n_x=0, 1, 2, ...$ and $n_y=0, 1, 2, ...$, independently. For the energy level $n=0$,
there is only a single eigenfunction -- the ground state -- $\phi_{0,0}$ [with $n_x=0=n_y$]. The next level
$n=1$ has two eigenfunctions: $\phi_{0,1}$ and $\phi_{1,0}$; therefore it is {\em degenerate}. A little
further thought quickly makes it clear that still higher energy states have even greater degeneracy [i.e.,
there are even more eigenfunctions sharing the same eigenenergies.] Here, the source of this degeneracy is
the rotational symmetry of the oscillator itself -- it is called {\em symmetry degeneracy}.

To grasp the role of the symmetry, it is convenient to write the Hamiltonian operator in the cylindrical
coordinates when the eigenfunction in the polar coordinates reads as $\psi(r,\phi)$. What becomes
immediately noticeable is that the (angular) coordinate $\phi$ does not appear except as a variable for
differentiation. When this happens in a classical Hamiltonian, we say it is {\em cyclic} in that coordinate
and the conjugate momentum (i.e., $p_{\phi}$) is a constant of motion. The quantal version of this statement
is that the Hamiltonian operator $\hat{H}$ commutes with the operator ($\hat{L_z}$) corresponding to that
momentum: i.e., [$\hat{H}, \hat{L_z}$]=0. It is easy enough to verify this relation, but less clear why this
is so. The reason that this happens in {\em this case} is that the system [i.e., the Hamiltonian] is
{\em invariant under rotation} --it is {\em isotropic}.

The degree of degeneracy $g_n$ can be calculated relatively easily [see, e.g., standard textbooks on quantum
mechanics]. For general $N$ dimensional (isotropic) harmonic oscillator characterized by quantum number $n$,
the formula for the degeneracy is given by $g_n=\binom {N+n-1}{n}$. This immediately reaffirms that in the
one-dimensional case each energy level corresponds to a {\em unique} quantum state and hence the system as
such stands as non-degenerate.

We have seen above how the symmetry leads to the degeneracy. If the situation were somehow changed to break
the symmetry, then this will {\em break} the degeneracy. This can be exemplified by considering, e.g.,
the anisotropic strengths of the harmonic oscillator along the x and y axes.

\subsection{The Wannier function and the Bloch function}

The Wannier function of a band is defined in terms of the Bloch function of the same band by
\begin{equation}
\chi(z-z_n)\,=\,\frac{1}{\sqrt{N}}\,\sum_{k}\,e^{-ikz_n}\,\phi_k(z),
\end{equation}
where $z_n$ is the spatial point in the lattice with period $d$. For the Bloch function
\begin{equation}
\phi_k(z)\,=\,e^{ikz}\,u(z)
\end{equation}
the Wannier function is
\begin{equation}
\chi(z-z_n)\,=\,\sqrt{N}\,u(z)\,\frac{\sin[\pi\,(z-z_n)/d]}{[\pi\,(z-z_n)/d]},
\end{equation}
where $u(z)=u(z+d)$ is the solution of the Schr\"{o}dinger equation [see Appendix A]
\begin{equation}
\left [-\frac{\hbar^2}{2m^*}\frac{d^2}{dz^2} + V(z)\right ]u(z)\,=\,\epsilon_0\, u(z)
\end{equation}
The foregoing diagnosis tells us how the kronig-Penney (KP) wave function $u(z)$, the Bloch function $\phi_k(z)$,
and the Wannier function $\chi(z-z_n)$ are very tightly interwoven with each other. Since the
characteristic eigenfunction in the present problem is defined in terms of these functions [see Eq. (1)], we
choose to study their graphic behavior. This is illustrated in Fig. 2. We believe it unnecessary to expand on the
obviously correct behavior of each function, since, as they say, {\em a picture is worth a thousand words}.

\begin{figure}[htbp]
\includegraphics*[width=8cm,height=12cm]{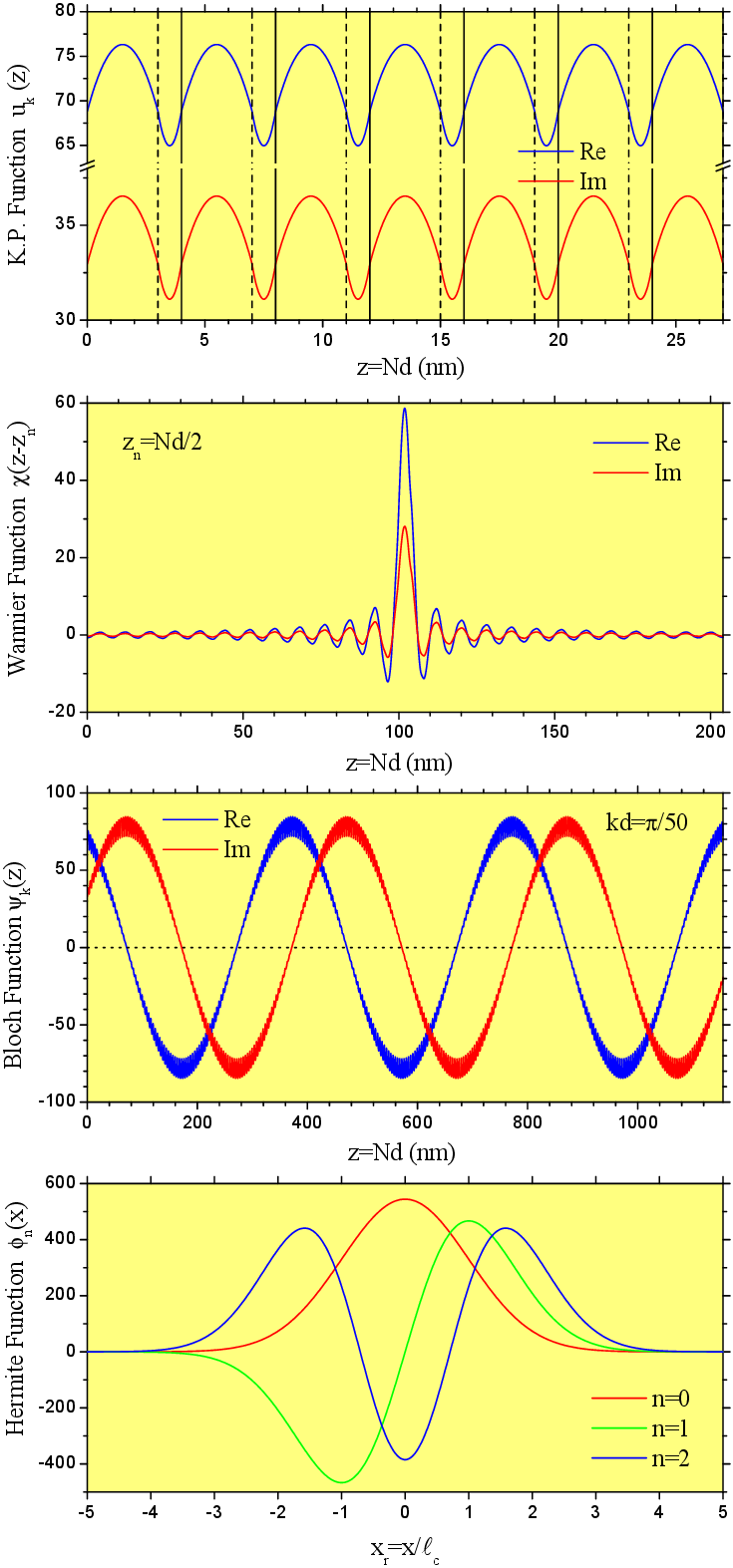}
\caption{(Color online) Graphic behavior of the KP, the Wannier, the Bloch, and the Hermite functions as used
in this work. The well (barrier) thickness is 3 (1) nm and $V_0=349.11$ meV.}
\label{fig2}
\end{figure}

\subsection{Miniband structure due to the BBC}

The introduction of the a superlattice potential perturbs the band structure of the host materials. The degree of
such perturbation depends very much on its amplitude and the periodicity. Since the superlattice period is usually
larger than the lattice constants of the (bulk) host constituents, the Brillouin zone is divided into minizones
giving rise to a series of narrow allowed bands separated by forbidden regions at the zone boundaries ($kd=n\pi$).
The allowed bands and forbidden regions are known, respectively, as minibands and minigaps in the language of
solid state physics. The band-gap engineering -- the process of controlling the band-gap by altering the
composition of the host components -- is a subject of paramount importance both for fundamental physics and device applications.

\begin{figure}[htbp]
\includegraphics*[width=8cm,height=11cm]{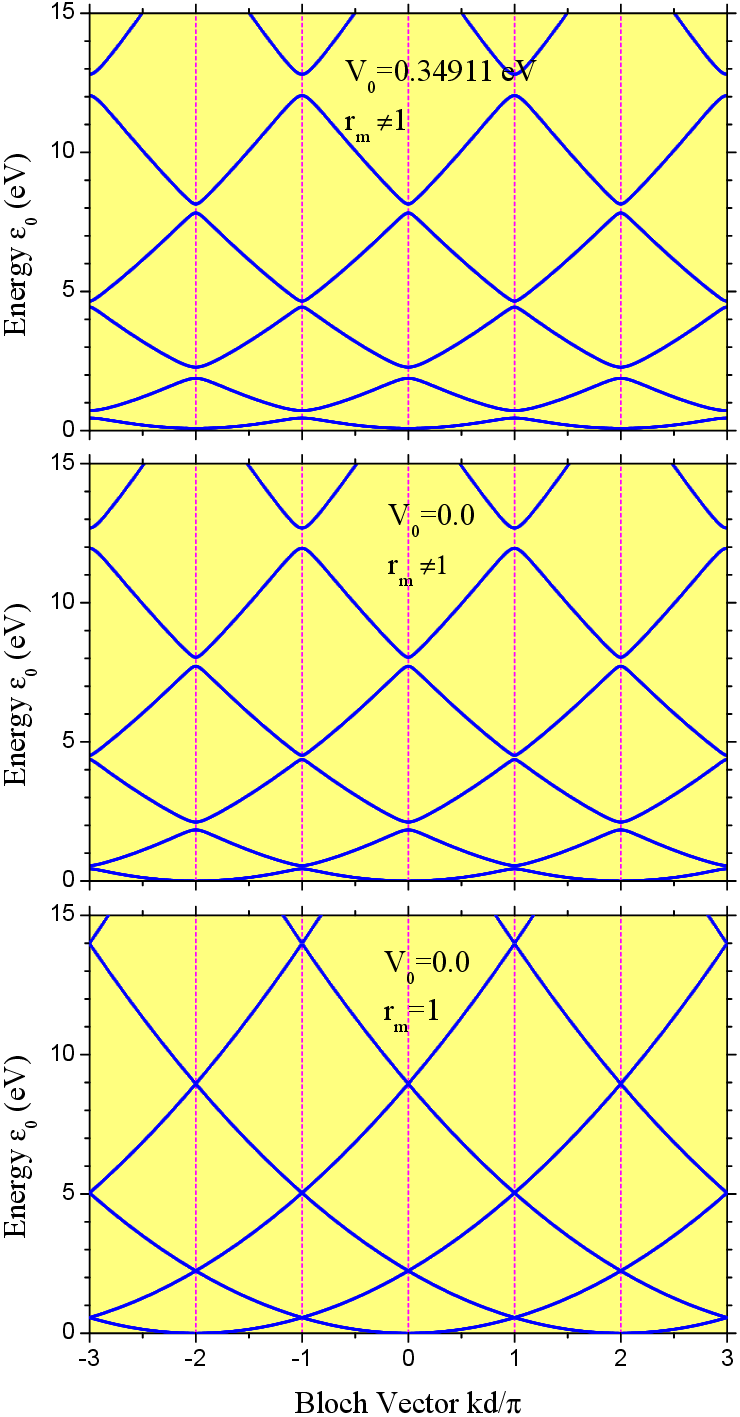}
\caption{(Color online) The miniband spectrum: the (KP) energy vs. the reduced Bloch vector. The top panel
represents the superstructure with a finite potential barrier height [$V_0=349.11$ meV] and the effective mass
ratio $r_m \ne 1$. The middle panel stands for the ``free particle" [$V_0=0$] with $r_m \ne 1$. The bottom panel
refers to the ``free particle" with $r_m =1$. We draw attention to the ``free" particle observing the gaps at the
zone boundaries ($kd=n\pi$) [see the middle panel].}
\label{fig3}
\end{figure}

It is known that a free particle (electron or hole) does not observe any gap in the band structure at the zone
boundaries. Yet, it turns out that in a superstructure with heterointerfaces, a ``free" electron ($V_0=0$) can
(and does) experience gaps at the zone boundaries provided that the dispersion relation is derived by employing
the proper effective-mass boundary conditions becoming known as the BBC. This is illustrated in Fig. 3 (see,
e.g., the middle panel). This implies that a ``free" electron virtually feels an {\em effective} potential
barrier due, in fact, to the dissimilar hosts (with different $m^*$) in the wells and the barriers. It is
noteworthy, however, that the magnitude of the gaps for the free particle (the middle panel) is relatively smaller
than that for the bound particle (top panel).


\begin{figure}[htbp]
\includegraphics*[width=8cm,height=9cm]{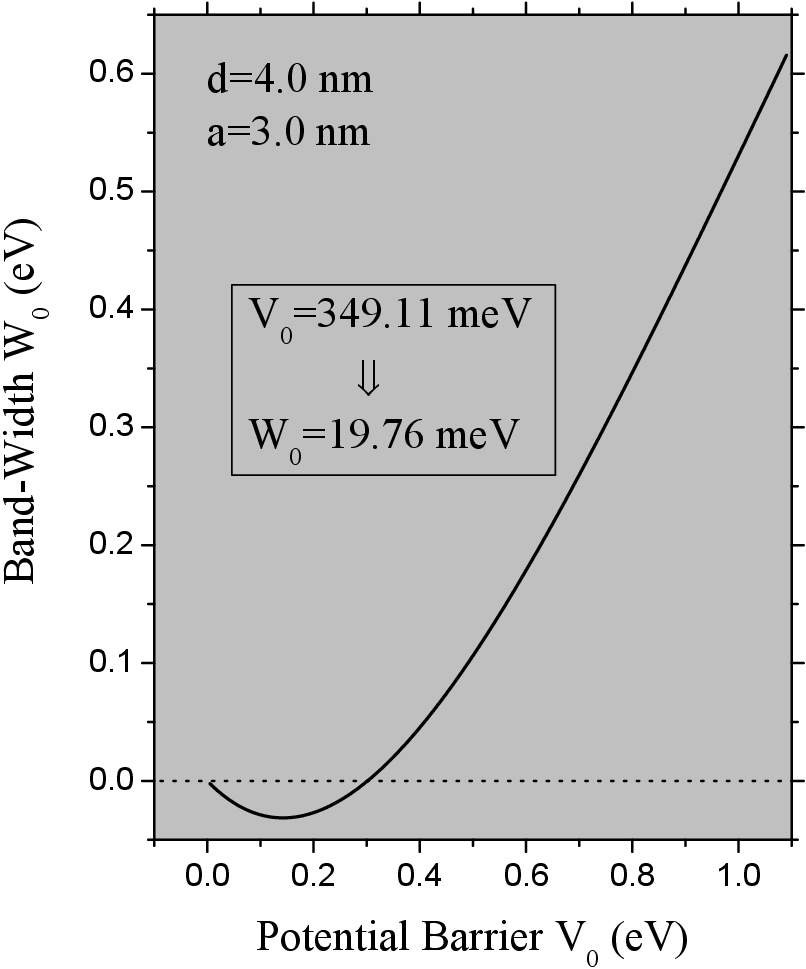}
\caption{The bandwidth of the (lowest) miniband as a function of potential barrier height $V_0$. The rest of the
parameters are as listed inside the picture.}
\label{fig4}
\end{figure}

\subsection{On the significance of the bandwidth}

A superlattice (periodic) potential gives rise to the miniband spectrum because of the quasi-free motion of the
charge carriers along the growth axis. The bandwidth of these minibands can be controlled by varying, e.g., the
composition of the superstructure, temperature, electric field, and magnetic field. At the same time, the
bandwidth can have significant control over numerous electronic, optical, and transport phenomena. This is
because the bandwidth is, in part, a measure of the electron dynamics along the superlattice axis. If the ratio
of the barrier height to the miniband bandwidth is much smaller than unity, electrons can roam almost freely and
influence the degree of tunneling. On the other hand, if this ratio is much greater than unity, electrons are
mostly confined within the quantum-dot islands. Since the bandwidth in the present problem governs the excitation
spectrum [see, e.g., Eq. (4)], we thought it worthwhile to study the bandwidth as function of the potential
barrier. This is shown in Fig. 4. It deserves mention that we choose to work with a bandwidth $W_0$ (=19.76 meV)
$<$ $\hbar \omega_{opt}$ ($\gtrsim 30$ meV) so as to avoid the optical phonon scattering. While it is not of
vital interest here, we notice that the bandwidth $W_0$ attains a negative value for the barrier height in the
range $0.00521 \lesssim V_0 \, (\rm eV) \lesssim 0.30099$. Energy bands with positive (negative) frequencies with
respect to the (chosen) zero of the energy scale refer to the positive (negative) bandwidths. For real signals,
the negative frequencies are always present and are always a mirror of the positive frequencies.

\subsection{On limiting the number of subbands}

Equation (24) is, in general, an $\infty \times \infty$ matrix until and unless we cut off the number of subbands
(and hence limit the electronic transitions) involved in the problem. First of all, we limit ourselves to the
lowest miniband (i.e., $t, t'\equiv 0$). Next, it is noteworthy that while experiments may report multiple subbands
occupied, theoretically it is not only extremely difficult but also (almost) impossible to compute the excitation
spectrum for the multiple-subband model. This is because the generalized dielectric function turns out to be a
matrix of the dimension of $\eta^2 \times \eta^2$, where $\eta$ is the number of subbands in the model. Handling
such enormous matrices (for a very large $\eta$) analytically is a {\em hard nut to crack} and if someone wants to
do it numerically then what one is ultimately left with is just a numerical simulation, which, generally, becomes
a matter out of taste for someone who likes to seek an understandable correspondence between the analytical and
the numerical results. For this reason, we choose to keep the complexity to a minimum and limit ourselves to a
two-subband model ($n, n', m, m' \equiv 1, 2$) with only the lowest one occupied. This is known to be quite a
reasonable assumption for these low-density, lower-dimensional (quantum) systems at lower temperatures where most
of the (electronic, optical, and transport) experiments are performed. This implies that the generalized dielectric
function is to be a $4 \times 4$ matrix. First, let us cast Eq. (24) in the following form.
\begin{equation}
\epsilon_{nn'mm'} (q; \omega)=\delta_{nm}\,\delta_{n'm'} - A_{nn'}(q, \omega)\,P_{nn'mm'}(q),
\end{equation}
where
\begin{equation}
A_{nn'}(q,\omega)=\Pi_{nn'}(k,k'=k+q;\omega)\,S_{00}(q),
\end{equation}
where $S_{00}(q)$ is just as defined in Eq. (20) [with $t, t'\equiv 0$] and
\begin{equation}
P_{nn'mm'}(q)=\int d\overline{r}\int d\overline{r}'\,
\phi^*_{n}(\overline{r})\,\phi^*_{n'}(\overline{r})\,
V_{ee}(q, \overline{r}-\overline{r}')\,
\phi^*_{m'}(\overline{r}')\,\phi_{m}(\overline{r}'),
\end{equation}
where
\begin{equation}
V_{ee}(q, \overline{r}-\overline{r}')=\frac{2e^2}{\epsilon_b}\,K_0(q\mid \overline{r}-\overline{r}'\mid)
\end{equation}
Now, the generalized dielectric function matrix in Eq. (29) takes the following form.
\begin{equation}
\tilde{\epsilon}(q,\omega)=
\left [
\begin{array}{cccc}
1-A_{11}\,P_{1111} \ & \ -A_{11}\,P_{1112} \ & \ -A_{11}\,P_{1121} \ & \ -A_{11}\,P_{1122} \\
-A_{12}\,P_{1211} \ & \ 1-A_{12}\,P_{1212} \ & \ -A_{12}\,P_{1221} \ & \ -A_{12}\,P_{1222} \\
-A_{21}\,P_{2111} \ & \ -A_{21}\,P_{2112} \ & \ 1-A_{21}\,P_{2121} \ & \ -A_{21}\,P_{2122} \\
-A_{22}\,P_{2211} \ & \ -A_{22}\,P_{2212} \ & \ -A_{22}\,P_{2221} \ & \ 1-A_{22}\,P_{2222} \\
\end{array}
\right ]
\end{equation}
Notice that $A_{22}=0$ because the second subband is unoccupied. Then, since the quasi-particle excitations
are given by $\mid \tilde{\epsilon}(q, \omega) \mid=0$, Eq. (33) finally yields
\begin{equation}
(1-A_{11}\,P_{1111})\,(1-B_{12}\,P_{1212}) - A_{11}\,B_{12}\,P^2_{1112}=0,
\end{equation}
where $B_{12}=A_{12}+A_{21}=\chi_{12}S_{00}$ (with $\chi_{12}=\Pi_{12}+\Pi_{21}$) is the intersubband
polarizability function that accounts for upward as well as downward transitions. This is the final equation
which has to be treated at the computational level. However, further simplification is possible depending on
the nature of the confining potential (see next).

\subsection{On the symmetry of the confining potential}

It is becoming known [1] that for a symmetric potential well, $P_{ijkl}$ (the Fourier-transformed Coulomb
interaction) is strictly zero provided that $i+j+k+l$ is an {\em odd} number. This is so because the
corresponding wave function is either symmetric or antisymmetric under space reflection. Since the
two-dimensional parabolic confining potential in the InAs islands is symmetric, $P_{1112}=0$ in Eq. (34).
This tells us that the intrasubband and intersubband modes represented, respectively, by the first and
second factors in the first term in Eq. (34) are decoupled, because the (second) coupling term is zero.
That $P_{1112}=0$ can be immediately verified just by rewriting Eq. (31) in the polar coordinates. Since
the subband index 0 (1) is allowed for the intrasubband (intersubband) excitation, we still need to conform
Eq. (34) such that the subscript $1\rightarrow 0$ and $2 \rightarrow 1$ for all practical purposes.

\subsection{The zero-temperature limit}

Since extremely low-temperatures are preferred for the experiments performed on the lower dimensional systems,
we choose to confine ourselves to the zero temperature limit. We think the temperature dependence of our results
would be significant only at $T \gtrsim 35$ K. In that situation, we can change the sum (over $k$) to an integral
by using the replacement rule and replace the Fermi distribution function with the Heaviside unit step function
such that
\begin{equation}
f(\epsilon)=\theta(\epsilon_F - \epsilon)=\left \{
\begin{array}{c}
1 \,\,\,\,\, {\rm if \,\,\,\,\, \epsilon_F > \epsilon}\\
0 \,\,\,\,\, {\rm if \,\,\,\,\, \epsilon_F < \epsilon}
\end{array}
\right . ,
\end{equation}
where $\epsilon_F$ is the Fermi energy in the problem. It is interesting to notice that in the zero-temperature
limit, we are allowed to go further to calculate analytically the polarizability functions $\Pi_{00}$ and
$\chi_{01}$. To this end, we first need to simplify the sum over $k$ [in, e.g., Eq. (14)]. This results in a
definite integral with the upper (lower) limit as $k_F$ ($-k_F$). Given the multiple definitions of that
integral (and the lack of a {\em single} definitive source) [45], we think it is more important to define here
that integral instead of writing the (lengthy) expressions of $\Pi_{00}$ and $\chi_{01}$. The integral reads as
\begin{eqnarray}
\int dx && \frac{1}{[a+b\cos(x)+c\sin(x)]} \nonumber\\
&&=\frac{2}{\sqrt{+a^2-b^2-c^2}}\,\tan^{-1}\left [\frac{c+(a-b)\,\tan(x/2)}{\sqrt{+a^2-b^2-c^2}}\right ],
\,\,\, {\rm  a^2 > b^2+c^2}\\
&&=\frac{-2}{\sqrt{-a^2+b^2+c^2}}\,\tanh^{-1}\left [\frac{c+(a-b)\,\tan(x/2)}{\sqrt{-a^2+b^2+c^2}}\right ],
\,\,\, {\rm a^2 < b^2+c^2; \mid y\mid < 1}\\
&&=\frac{-2}{\sqrt{-a^2+b^2+c^2}}\,\coth^{-1}\left [\frac{c+(a-b)\,\tan(x/2)}{\sqrt{-a^2+b^2+c^2}}\right ],
\,\,\, {\rm a^2 < b^2+c^2; \mid y\mid > 1}
\end{eqnarray}
where $y=$ [the whole quantity within the square brackets in the respective equality]. The symbols $a$, $b$, $c$
are not to be confused with the well/barrier widths. Although one can analytically convert one form into the
other, the computer does not know this. What follows next is to express the hyperbolic functions in terms of the
logarithmic ones in order to obtain the manageable forms of $\Pi_{00}$ and $\chi_{01}$. We analyze the resulting
forms of $\Pi_{00}$ and $\chi_{01}$ in the long wavelength limit to obtain
\begin{equation}
\Pi_{00}=\frac{n_{_{1D}}\, q^2 }{m^*_e \omega^2} + O(q^4),
\end{equation}
where $m^*_e\,[=\hbar^2/(W_0 d^2)]$ stands for an {\em effective} mass in the problem, and
\begin{equation}
\chi_{01}=\frac{2\,n_{_{1D}}\, \epsilon_{21}}{(\hbar \omega)^2-\epsilon_{21}^2} + O(q),
\end{equation}
where $\epsilon_{21}=\hbar \omega_0$ is the subband spacing. It is interesting to note that the long wavelength
forms of the polarizability functions derived in Eqs. (37) and (38) are independent of the dimensionality of the
system and have the same forms for higher dimensions as well [1].


\subsection{The Density of states and the Fermi energy}

An analytical treatment requires that we start with Eq. (4). We derive the following expression for computing self-consistently
\begin{equation}
g(\epsilon)=\frac{2}{\pi d}\,\sum_{n}\,
       \left [W_h^2-(\epsilon-\epsilon_n)^2\right ]^{-1/2}\,
\theta \left [W_h^2-(\epsilon-\epsilon_n)^2\right ],
\end{equation}
the density of states (DOS), and
\begin{equation}
n_{1D}=\frac{2}{\pi d}\,\sum_{n}\,
\cos^{-1}\left [-\frac{1}{W_h}\,(\epsilon_F - \epsilon_n)\right ]\,
   \theta\left [-\frac{1}{W_h}\,(\epsilon_F - \epsilon_n)\right ],
\end{equation}
the Fermi energy. Here $W_h=W_0/2$, $\epsilon_n  =(n+1)\hbar \omega_0 + \epsilon_0$, and $n_{1D}$ is the linear
charge density of the electrons (i.e., the number of electrons per unit length). Note that $\epsilon_0$ is the
energy of the lowest miniband at the zone center and can be neglected with no loss of generality (because it
affects neither the intra- nor the inter-subband excitations).

\begin{figure}[htbp]
\includegraphics*[width=8cm,height=9cm]{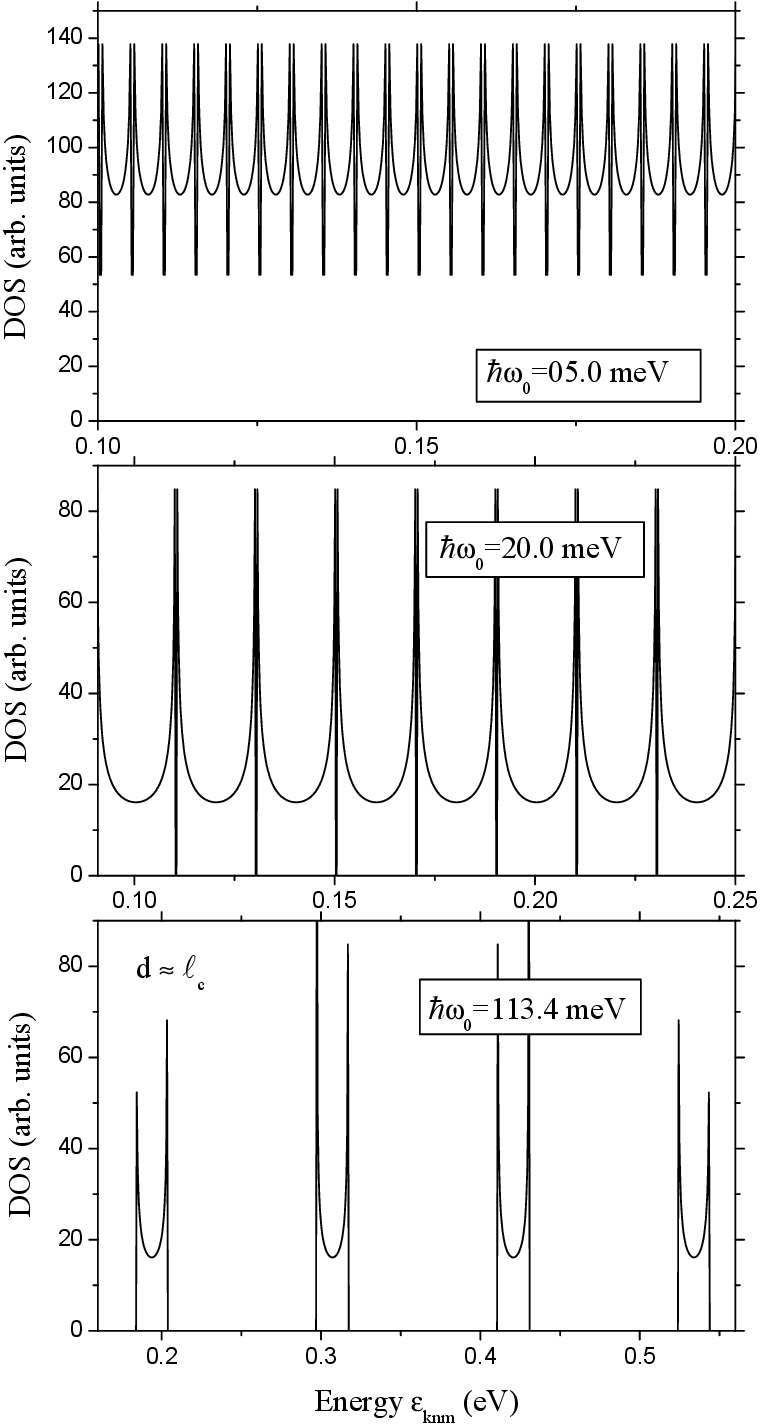}
\caption{The density of states vs. the excitation energy for the resultant quantum wire. The confinement potential
(or subband spacing) is defined as $\hbar \omega_0=5.0$ meV (top panel), 20.0 meV (middle panel), and 113.4 meV
(bottom panel). The subband spacing in the bottom panel was designed to have $\ell_c \simeq d$. The band-width is $W_0=19.76$
meV.}
\label{fig5}
\end{figure}

Figure 5 depicts the results on the density of states versus the excitation energy. We observe that as the
confinement increases the DOS structure (in the top panel) is pushed downwards to take finally the shape shown in
the middle panel where all the spikes visible in the upper panel start from the origin. With still stronger
confinement, the structure in the DOS starts splitting in energy as seen in the bottom panel, where the resultant
system favors the decoupled InAs islands. In other words, very strong confinement seems to prevent the forming of
the quantum wire due, virtually, to little quasi-free motion along the growth direction. This also reaffirms the
fact that the DOS is very much a dimensionality dependent property of the system and the role dimensions play is
evident from the units of the DOS: $g(\epsilon) \propto \epsilon^{1/2}$ (in 3D), $g(\epsilon) \propto \epsilon^{0}$
(in 2D), $g(\epsilon) \propto \epsilon^{-1/2}$ (in 1D). In three-dimensionally confined quantum dots the DOS is
known to be $\delta-$function-like leading to the well-argued vanishing of the thermal broadening. We had purposely
designed the confinement in the bottom panel in order to have a competition between the period ($d$) and the
characteristic length of the harmonic oscillator ($\ell_c$) [i.e., to have $d \simeq \ell_c$]. However, we did not
observe any specific effect due to this condition.

\begin{figure}[htbp]
\includegraphics*[width=8cm,height=9cm]{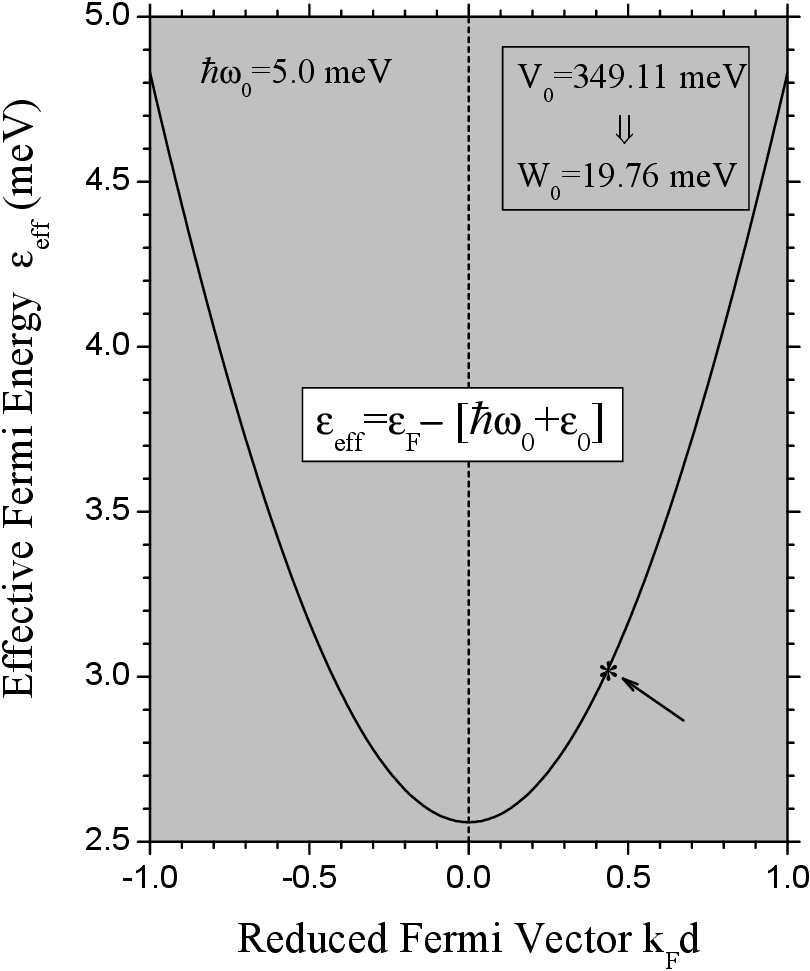}
\caption{The Fermi energy vs. the reduced Fermi vector for the resultant quantum wire. We purposely made this
choice to have $k_F d$ on the abscissa -- it could have been simply the charge density $n_{1D}$. The subband
spacing $\hbar \omega_0=5.0$ meV and the band-width $W_0=19.76$ meV. The $\star$ on the Fermi energy curve
corresponds to $\epsilon_{eff}=3.0296$ meV.}
\label{fig6}
\end{figure}

Figure 6 illustrates the (effective) Fermi energy as a function of the dimensionless Fermi vector ($k_F d$).
Since the Fermi vector is related to the 1D charge density by means of $k_F=(\pi/2) n_{1D}$, one could choose
to plot the Fermi energy versus the charge density. However, we made this choice so as to feel free in
choosing, e.g., $n_{1D}$ and $d$ simultaneously. The Fermi energy curve turns out to be symmetric because the
Fermi vector (or the Fermi number) happens to be inside the argument of a cosine term in the single-particle
energy. The $\star$ on the Fermi energy curve refers to the effective Fermi energy $\epsilon_{eff}=3.0296$ meV
-- a value that corresponds to $k_F d=0.4398$. This implies that $n_{1D}=0.7 \times 10^{6}$ cm$^{-1}$, for a
period of $d=4.0$ nm. We will see in the next section that our choice of the subband spacing
$\hbar \omega_0=5.0$ meV in a two-subband model (with only the lowest one occupied) is justifiable since the
Fermi energy, for this set of parameters, lies at $\epsilon_{eff}=3.0296$ meV (i.e., $0 < \epsilon_{eff}
<\hbar \omega_0$). The chemical potential at zero temperature is equal to the Fermi energy. The knowledge of the
variation of the Fermi energy is of cardinal importance to the understanding of (almost all) electronic, optical,
and transport phenomena in a quantal system. This is true to the extent that the transport properties (such as
conductance, resistance, ..etc.) are {\em reflections} of the electron dynamics at/near the Fermi surface in the
system. The fascinating thing about the Fermi surface is that you can tailor it before it tailors the rest.

\subsection{On the behavior of $S_{00}(q)$}

In this subsection, we would like to see how the factor $S_{00}(q)$, which virtually governs the quasi-free motion
along the growth direction, behaves as the momentum transfer $q$ and/or the well-width $a$ varies. This is shown
in Fig. 7, where we plot $S_{00}(q)$ as a function of momentum transfer $q/k_F$ (in left panel) and as a function
of well-width $a$ (in the right panel). As seen in the left panel, $S_{00}$ starts with a finite value at the zone
center and decreases monotonously so as to form a half inverted parabola centered at its starting value. When we
fix the momentum transfer and vary the well-width, the factor $S_{00}$ again starts from a nonzero value at
$a \simeq 2.78$ nm and tends to decrease in such a manner as to make a half parabola centered at $a \simeq 8.2$ nm.
Repeating this procedure for higher values of the barrier widths $b$ reveals that while the factor $S_{00}$
increases with increasing $b$, it exhibits a similar behavior as for the lower values of $b$. In addition, the
factor $S_{00}$ seems to saturate at $a \simeq 8.47$ nm for higher values of $b$. Since $S_{00}$ only contributes to
the collective excitations, single-particle excitations remain unaffected by its very existence in the formulation.


\begin{figure}[htbp]
\includegraphics*[width=8cm,height=9cm]{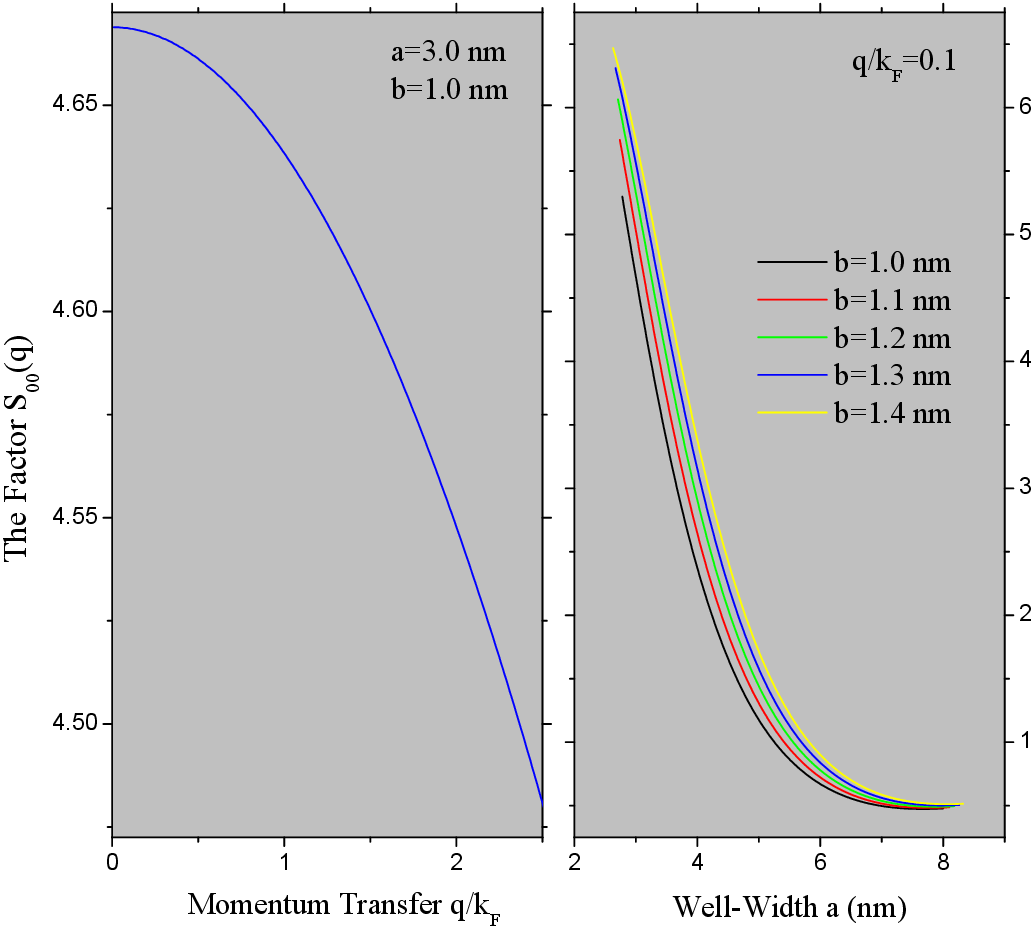}
\caption{(Color online) The factor $S_{00}$ vs. reduced momentum transfer $q/k_F$ (left panel) and vs. well-width
$a$ (right panel). The parameters involved in the computation are listed inside the picture.}
\label{fig7}
\end{figure}

\section{Illustrative Examples on the plasmon excitations}

For the illustrative numerical examples, we focus on the InAs/GaAs system like the one in the original experiment
[3]. The material parameters used are: effective mass $m^*=0.042 m_{_0}$ ($0.067 m_{_0}$) and the background
dielectric constant $\epsilon_{_b}=13.9$ ($12.8$) for the InAs (GaAs). We use the potential barrier of height
$V_0=349.11$ meV that produces the band-width (of the lowest miniband) to be $W_0=19.76$ meV, in compliance with
Sakaki [2] so as to be beyond the regime of the optical phonon scattering. Further, we take the subband spacing $\hbar\omega_{_0}=5.0$ meV, the self-consistently determined effective Fermi energy $\epsilon_{eff}=3.0296$ meV
for a 1D charge density $n_{1D}=0.7 \times10^6$ cm$^{-1}$, and the effective confinement width of the parabolic
potential well, estimated as the FWHM of the extent of the Hermite function,
$w_{eff}=2\sqrt{2 \ln (2)}\sqrt{n+1}\,\ell_{_c}=44.857$ nm. The layer widths will be specified case-wise while
discussing the results. Notice that the Fermi energy $\epsilon_F$ varies when the charge density ($n_{1D}$) or the
confining potential ($\hbar \omega_{0}$) is varied. Thus we aim at exploring the single-particle and collective
excitations in a quantum wire made up of vertically stacked quantum dots in the absence of an applied magnetic
field at T=0 K in a two-subband model within the full RPA. The case of a nonzero (finite) magnetic field is
deferred to a future publication.

\subsection{The excitation spectra}

Figure 8 illustrates the full excitation spectrum composed of single-particle and collective (plasmon) excitations
for a quantum wire plotted as the energy $\hbar \omega$ vs. the reduced momentum transfer $q/k_F$, for the given
values of the charge density $n_{1D}$, subband spacing $\hbar\omega_0$, and the layer thicknesses $a$ and $b$. Note
that this is the case when the well width $a$ is greater than the barrier width $b$. The light (dark) shaded region
refers, respectively, to the intrasubband (intersubband) single-particle excitations (SPE). The bold solid curve
marked as $\Omega_{00}$ ($\Omega_{10}$) is the intrasubband (intersubband) collective (plasmon) excitation (CPE).
The dashed curves inside and in the close vicinity of the upper edge of the SPE are also the CPE, but these modes
are Landau damped and we will not be discussing them further. The intrasubband CPE starts from the origin and is
not seen to merge with the upper edge of the intrasubband SPE until in the short wavelength limit. In that sense it
is a bonafide, long-lived plasmon which should be easily observed in, for example, the resonant Raman scattering.
The intersubband CPE starts at ($q/k_F=0$, $\hbar\omega=12.926$ meV), attains a minimum at ($q/k_F=0.61$,
$\hbar\omega=7.73$ meV), and then rises up to propagate very closely to the upper edge of the intersubband SPE.
However, even the instersubband CPE is not observed to merge entirely with the upper edge of the intersubband
SPE. Therefore, it also remains free from Landau damping until very large propagation vector. It is not difficult
to prove (analytically) why the energy of the lower branch of the intrasubband SPE goes to zero at $q=2k_F$, why
the lower branch of the intersubband SPE exhibits its {\em minimum} at $q=k_F$, and why the intersubband SPE starts
at the subband spacing ($\hbar\omega=5.0$ meV) at $q=0$. The most interesting aspects of this excitation spectrum
are: (i) both genuine collective excitations which are free from Landau damping, (ii) the existence of the
intersubband CPE which changes the sign of its group velocity before tending to merge with the respective SPE, and
(iii) the overlap of the intrasubband and intersubband SPE as compared to the Figs. 9 and 10 below. The ratio of
the intersubband resonance ($\omega_*$) to the subband spacing ($\omega_0$) at $q=0$ is found to be $\omega_*/\omega_0=2.5852$. Later we remark on this energy shift of the intersubband CPE with respect to the
corresponding SPE.

\begin{figure}[htbp]
\includegraphics*[width=8cm,height=9cm]{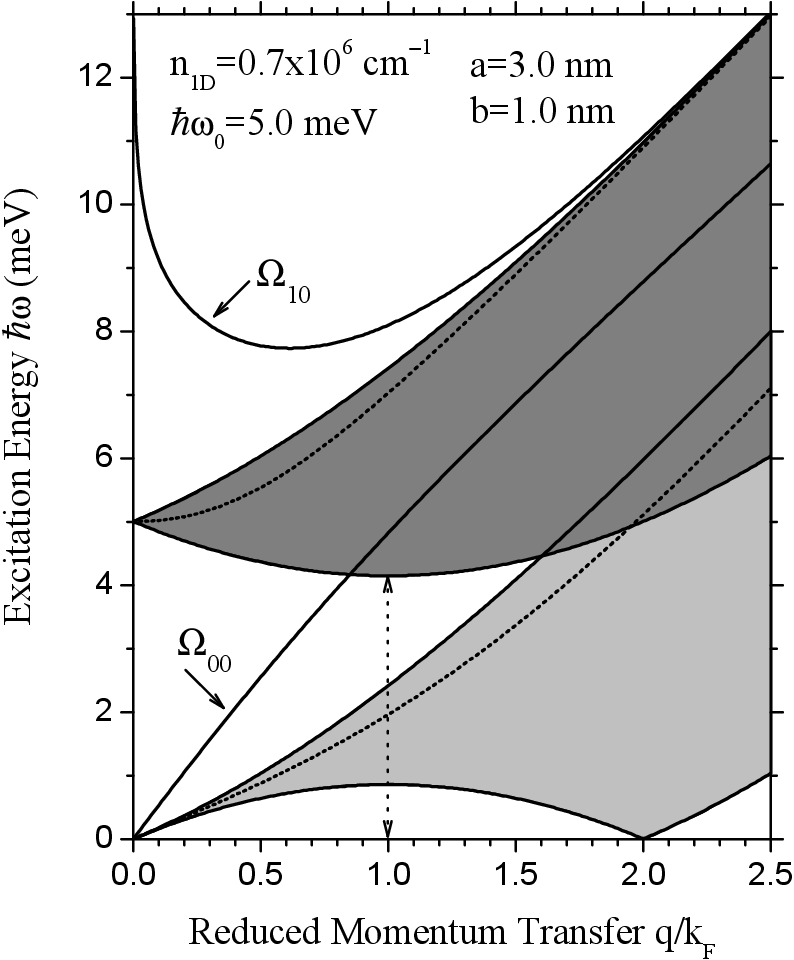}
\caption{The excitation spectrum of a quantum wire within a two-subband model where the energy $\hbar \omega$ is
plotted as a function of the reduced momentum transfer $q/k_F$. The light (dark) shaded region refers to the
intrasubband (intersubband) SPE associated with the lowest occupied (first excited) subband at $T=0$. The bold
lower (upper) curve represents the intrasubband (intersubband) CPE. The dashed curves inside the SPE are the
respective Landau damped CPE with no life-time at all. The vertical double-headed arrow stresses the minimum of
the intersubband SPE at $q=k_F$. We call attention to the intersubband CPE which changes the sign of its group
velocity before tending to merge with the respective SPE. The relevant parameters are as listed inside the picture.
It is noteworthy that the $a>b$.}
\label{fig8}
\end{figure}

Figure 9 shows the full excitation spectrum made up of single-particle and collective (plasmon) excitations for a
quantum wire plotted as the energy $\hbar \omega$ vs. the reduced momentum transfer $q/k_F$ just as in Fig. 8, but
for the case where the well and the barrier widths are equal (i.e., $a=b$). It should be pointed out that we
purposely choose minimum possible widths for both the well and the barrier so that the full excitation spectrum
could emerge. The importance of the variation in the layer thicknesses is remarkable. The whole spectrum
has considerably decreased in energy as compared to that in Fig. 8. There is a wide energy gap between the
intersubband and the intrasubband excitations. Apart from the fact that it now starts at lower energy
($\hbar \omega = 6.7094$ meV), the intersubband collective (plasmon) excitation has now smoothed out its wide dip
present in Fig. 8. In addition, the intersubband CPE is now seen to merge with the upper edge of the corresponding
SPE at $q/k_F \simeq 2.17$ and hence becomes Landau-damped thereafter. The ratio $\omega_*/\omega_0$ is now defined
as  $\omega_*/\omega_0=1.3418$. The rest of the discussion related with Fig. 8 is still valid.

\begin{figure}[htbp]
\includegraphics*[width=8cm,height=9cm]{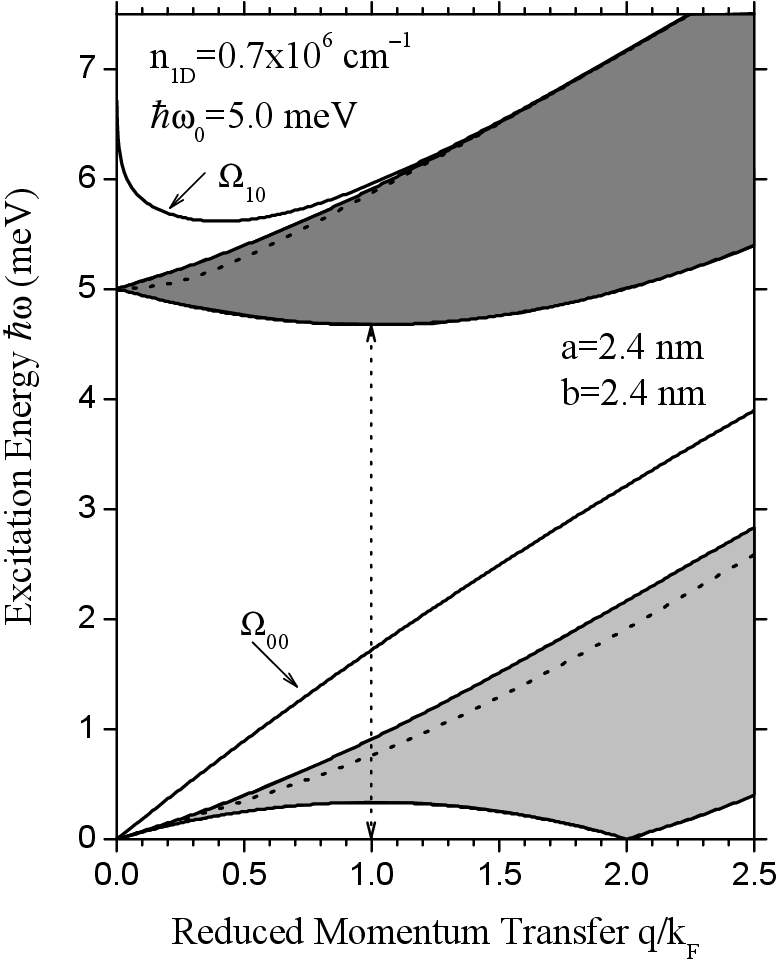}
\caption{The same as in Fig. 8, but for the case where the well and the barrier widths are equal (i.e., $a=b$). We
call attention to lowering of the whole spectrum in energy and smoothing out of the big dip in the intersubband CPE
in Fig. 8. The relevant parameters are as listed inside the picture.}
\label{fig9}
\end{figure}

Figure 10 depicts the full excitation spectrum comprised of single-particle and collective (plasmon) excitations for
a quantum wire plotted as the energy $\hbar \omega$ vs. the reduced momentum transfer $q/k_F$ just as in Fig. 8, but
for the case where the well and the barrier widths are such that $a<b$. Again while intending to keep the barrier
width greater than the well width, we kept to choose the minimum $a$ and $b$. We observe further lowering of
the whole spectrum in energy and a wider gap between the intersubband and intrasubband excitations as compared to
Fig. 9. It is, however, interesting to note that the intersubband CPE does not merge with the upper branch of the
respective SPE and remains free from Landau damping, unlike Fig. 9. The ratio $\omega_*/\omega_0$ is now given by $\omega_*/\omega_0=1.0427$. The drastic differences observed in the excitation spectrum in Figs. 9 and 10 as compared
to Fig. 8 can be intuitively understood as follows. The variation in the layer thicknesses in Figs. 9 and 10 have
virtually resulted in reduction of the well width as compared to the barrier width. This implies that the wavelength
of the respective excitations arising due to the charge carriers available in the wells has been implicitly enhanced
and consequently the frequency (or the energy) lowered.

\begin{figure}[htbp]
\includegraphics*[width=8cm,height=9cm]{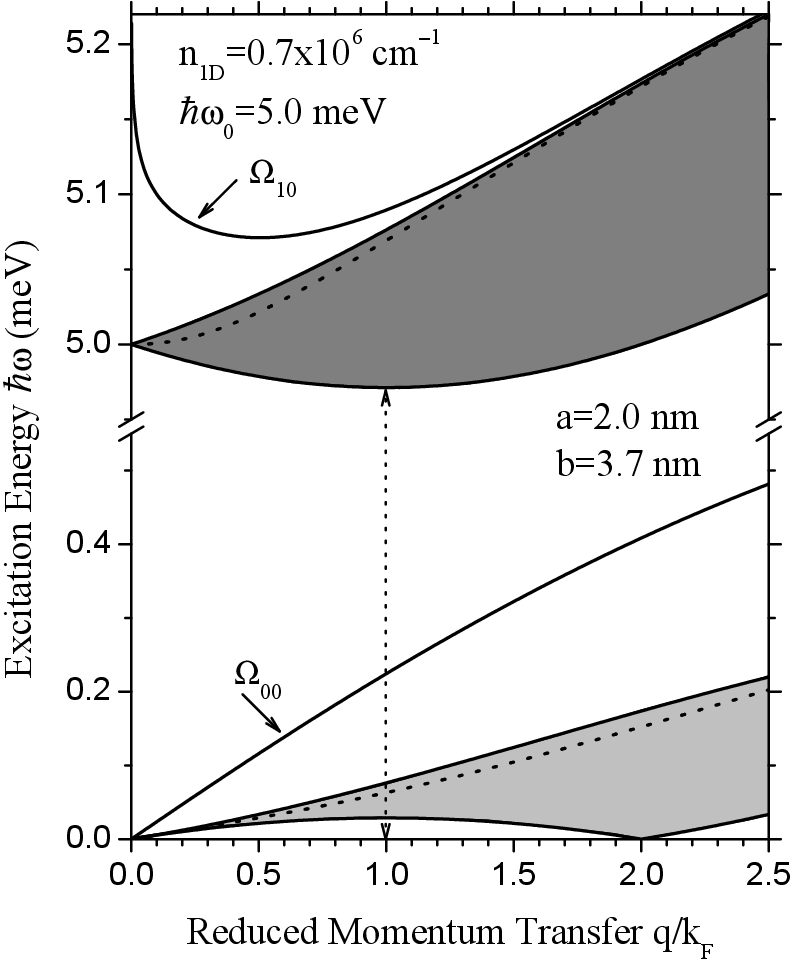}
\caption{The same as in Fig. 8, but for the case where the well and the barrier widths are again unequal in a manner
such that now $a<b$. We call attention to still further lowering of the whole spectrum in energy and wider gap
between the intersubband and intrasubband excitations as compared to Fig. 9. Notice the scale break on the energy axis.
The relevant parameters are as listed inside the picture.}
\label{fig10}
\end{figure}

We think that the discussions of the excitation spectrum in Figs. $8-10$ will remain incomplete until we shed some
light on the energy shift of the intersubband CPE with respect to the corresponding SPE at $q=0$. At the momentum
transfer $q=0$, the intersubband CPE generally starts at higher energy than the intersubband SPE (or the subband
spacing). This shift of the intersubband resonance ($\hbar\omega_*$) to an energy significantly higher than the
subband spacing ($\hbar\omega_{_0}$) is a manifestation of the many-body effects such as depolarization and
excitonic shifts [1]. We assume that the depolarization effects are dominant and thus have $\omega_*=(\omega_{_0}^2+\omega_d^2)^{1/2}$, where $\omega_d$ is the depolarization frequency. In the absence of a
desired quantal model, an upper bound on the depolarization shift can be obtained classically to yield
$\omega_d = \sqrt{8\,\pi \, e^2\, n_{1D}/(\tilde{\epsilon}\,m^*\,w_{eff}^2)}$.
With our input parameters, this requires an effective (background dielectric) constant
$\tilde{\epsilon}=1.11\,\epsilon_b$, $7.82\,\epsilon_b$, and  $71.64\,\epsilon_b$, respectively, in Figs. 8, 9, and
10. This signals the importance of the screening effects in such quantum wires. A large depolarization shift such as,
e.g., the one in Fig. 8 is, generally, not compensated by the excitonic shift (only reduced roughly by $20\%$). The
larger the ratio $\tilde{\epsilon}/\epsilon_b$ the stronger the screening effects and the better the chances of such compensation. In that sense the depolarization shift in Fig. 10 can be accounted for better than in Fig. 9 (and much
better than in Fig. 8).

\subsection{Influence of the layer thicknesses}

Since the well and the barrier widths were seen to affect the excitation spectrum strikingly, we thought it worthwhile
to study exclusively their influence on the collective excitations. Figure 11 illustrates the intrasubband CPE
plotted as a function of the barrier width $b$ for a given momentum transfer $q/k_F=0.05$ and for several values of
the well width $a$. First of all, it is noteworthy that for $b=0$ where the system reduces to a single (homogeneous)
quantum wire, the plasmon energy remains the same for all the cases as is intuitively expected. We observe that except
for a small rise in energy for $b \lesssim 0.5$ nm where the coupling is strong, the plasmon energy decreases with
increasing barrier width. For smaller $b$, the larger the $a$, the higher the plasmon energy, but, for larger $b$,
this trend is reversed. This behavior can also be interpreted in terms of the tunneling strength since the smaller
(larger) $b$ indicates the stronger (weaker) tunneling.

\begin{figure}[htbp]
\includegraphics*[width=8cm,height=9.0cm]{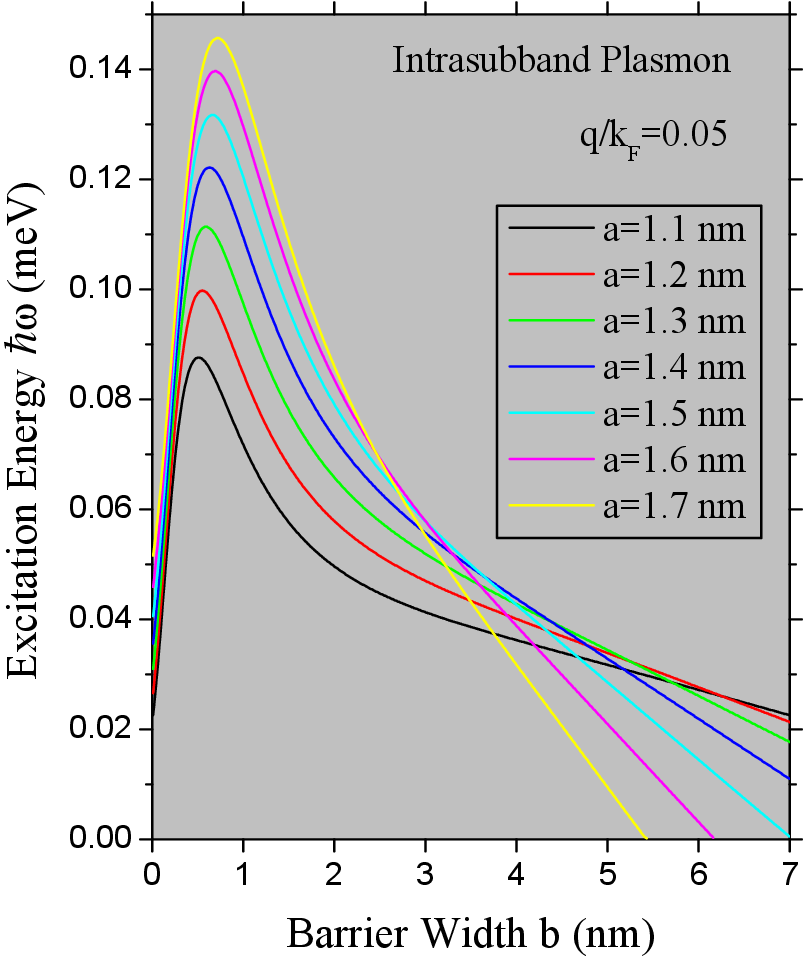}
\caption{(Color online) Intrasubband plasmon dispersion vs. the barrier width for several values of the well width.
The momentum transfer is fixed as $q/k_F=0.05$. The other parameters are the same as in Fig. 8.}
\label{fig11}
\end{figure}

Figure 12 shows the intrasubband plasmon energy as a function of the well width for a given value of momentum transfer
$q/k_F=0.05$ and for several values of barrier width. We separate the figure into two panels for the reasons of
clarity: the left (right) panel covers the range $0.0 < a \,(\rm nm) <3.0$ ($2.5 < a \, (\rm nm)<9.0$). It is
interesting to notice that the plasmon originates with infinitesimally small (but nonzero) energy at very small value
of $a$. This is what we intuitively expect. For $0.0 < a \,(\rm nm)\lesssim 2.75$, we observe that the smaller the $b$,
the higher the plasmon energy. However, as we approach the range $2.75\lesssim a \,(\rm nm) <9.0$ the preceding trend
is reversed. Notice that now it is much less because of the variation in the coupling (or the tunneling) strength but
more because of a different aspect inherent to the tight-binding approximation (TBA).
As the well width becomes larger and larger the Wannier functions gradually lose their strength. As a consequence, the coupling within the neighboring wells weakens and this leads to the reversal of the trend in the latter range of the
well width.

\begin{figure}[htbp]
\includegraphics*[width=8cm,height=9.0cm]{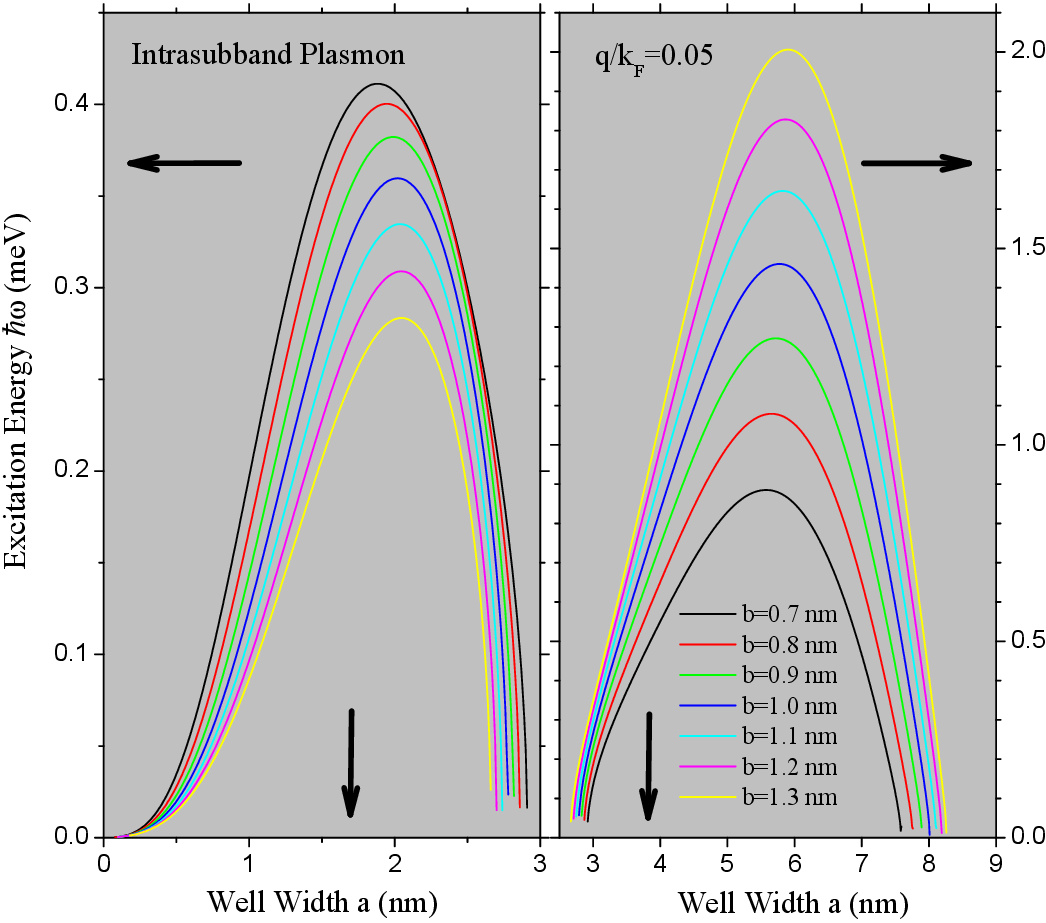}
\caption{(Color online) Intrasubband plasmon dispersion vs. the well width for several values of the barrier width.
The momentum transfer is fixed as $q/k_F=0.05$. We call attention to the difference in the energy and the well-width
scales between the two panels which were divided for the sake of clarity. The other parameters are the same as in
Fig. 8.}
\label{fig12}
\end{figure}

Figure 13 depicts the intersubband plasmon energy as a function of the barrier width for a given value of the momentum transfer $q/k_F=0.1$ and for several values of the well width. Before we discuss this and the next figure, we need
to consider an another issue: the existence of intersubband excitations requires the wells to be of moderate width;  extremely thin wells cannot excite the intersubband excitations. Given this, we see that the plasmon energy
obeys this rule in the range $0.0 < b \,(\rm nm)\lesssim 1.34$: the larger the $a$, the lower the plasmon energy. This
is again due to the characteristics of the Wannier functions as discussed above. For $b > 1.34$ nm, the trend is
reversed: i.e., the smaller the $a$ the lower the plasmon energy.

\begin{figure}[htbp]
\includegraphics*[width=8cm,height=9.0cm]{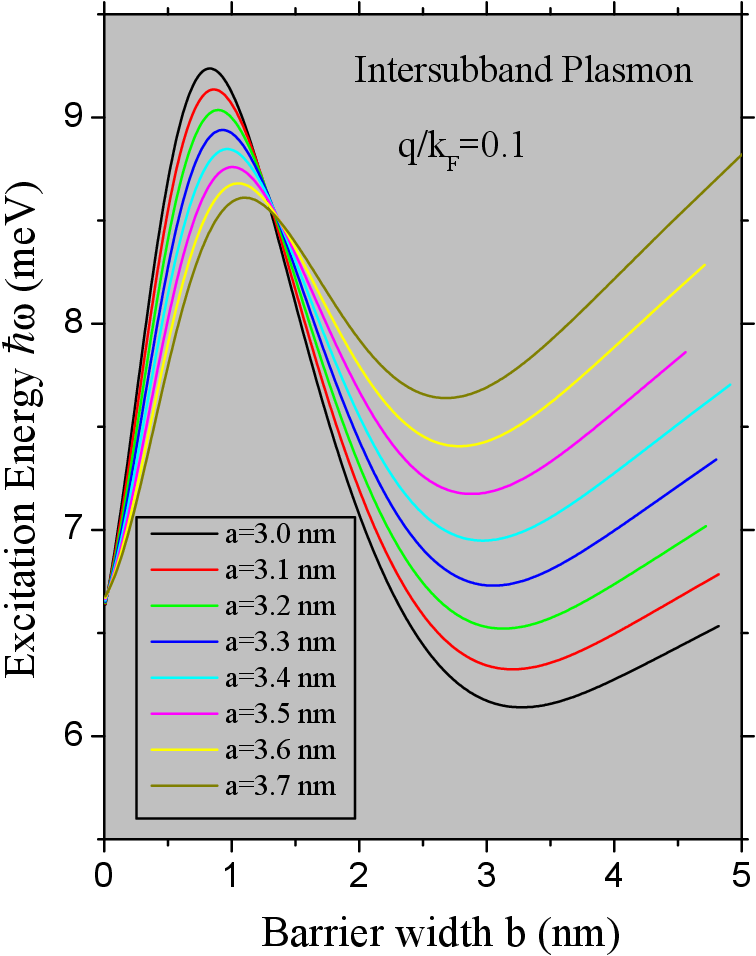}
\caption{(Color online) Intersubband plasmon dispersion vs. the barrier width for several values of the well width.
The momentum transfer is fixed as $q/k_F=0.1$. The other parameters are the same as in Fig. 8.}
\label{fig13}
\end{figure}

Figure 14 displays the intersubband plasmon energy as a function of the well width for a given value of the momentum
transfer $q/k_F=0.1$ and for several values of the barrier width. The discussion of this figure is very nearly the
same as that of Fig. 12, except for the fact that here the focal point [where all the modes roughly cross each other]
(at $a\simeq 4.0$) is rather sharper than it was in Fig. 12. We observe at a glance that the plasmon dispersion, in
the range $0.0 < a \,(\rm nm) \lesssim 4.0$, obeys this rule: the lower the $b$, the higher the plasmon energy. This
trend is, however, reversed after the focal point (i.e., for $a \gtrsim 4.0$ nm). Again, this is explicable in terms
of the Wannier function and its behavior as a function of the well width.


\begin{figure}[htbp]
\includegraphics*[width=8cm,height=9.0cm]{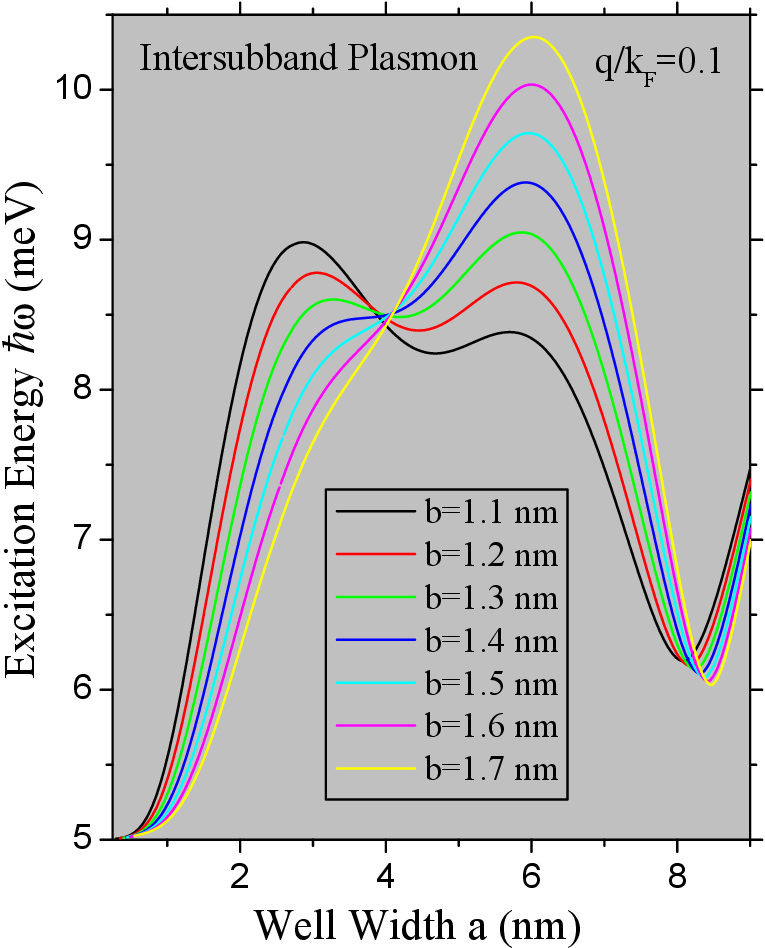}
\caption{(Color online) Intersubband plasmon dispersion vs. the well width for several values of the barrier width.
The momentum transfer is fixed as $q/k_F=0.1$. The other parameters are the same as in Fig. 8.}
\label{fig14}
\end{figure}

\subsection{The dependence on the charge density}

Figure 15 shows the dependence of the collective (plasmon) excitations on the charge density of the resultant quantum
wire system made up of the VSQD. It is noteworthy that we have plotted only the bonafide intrasubband and intersubband
CPE which remain Landau-undamped until a very large propagation vector before merging with the respective SPE. The same-colored upper (lower) mode is the intersubband (intrasubband) plasmon designated as $\Omega_{10}$ ($\Omega_{00}$)
in Fig. 8. Notice that the layer thicknesses are kept the same as in Fig. 8. In addition,  we have purposely avoided
to plot the respective single-particle continua in order not to make a mess in the picture. What we, generally, observe
in this figure is what we intuitively expect: the energy of the plasmon excitations increases with increasing charge
density. The literature is a live witness that this remark remains true irrespective of the size and dimension of the
system and we are not aware of any exception to this rule. It is, however, interesting to note that the sharp dip in
the intersubband CPE (where the group velocity changes the sign from negative to positive) is seen to be smoothed out
with the lowering of the charge density. Another interesting effect the variation of the charge density can have on
the excitation spectrum is that the single-particle continua (of intra- and inter-subband excitations) do not overlap
for $n_{1D}\lesssim 0.6\times10^{6}$ cm$^{-1}$.

\begin{figure}[htbp]
\includegraphics*[width=8cm,height=9cm]{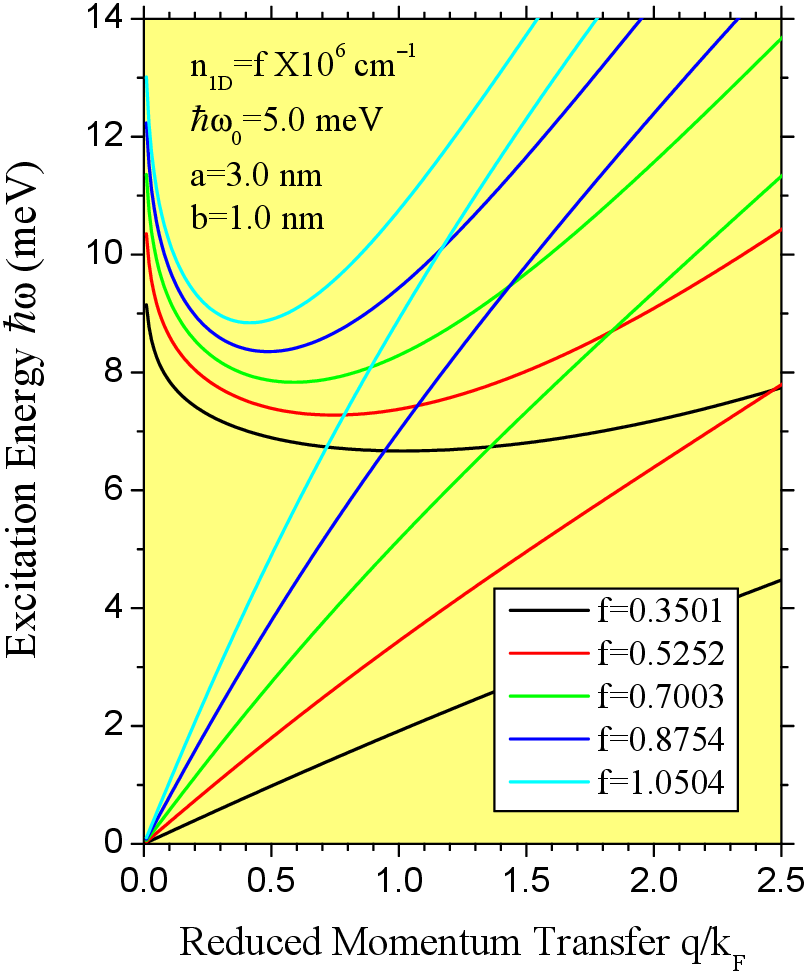}
\caption{(Color online) The collective (plasmon) excitation spectrum within a two-subband model where the energy
$\hbar \omega$ is plotted as a function of the reduced momentum transfer $q/k_F$, for several values of the charge
density $n_{1D}$. The other parameters are as listed inside the picture. Notice that we have plotted only the
bonafide intrasubband and intersubband plasmons which remain Landau-undamped until a very large propagation vector.}
\label{fig15}
\end{figure}

\subsection{On the inverse dielectric functions}

As mentioned before, the study of inverse dielectric function helps us substantiate the plasmon excitations generally
searched through the zeros of the dielectric function. This is what we should anticipate because the zeros of the DF
and the poles of the IDF must yield exactly identical results. There is an advantage of the latter over the former. A  careful analysis of the IDF also reflects over the longitudinal (transverse or Hall) resistance in such quantum
systems. For instance, the longitudinal (Hall) resistance $\rho_{zz}$ ($\rho_{yz}$) is determined by the imaginary
(real) part of the IDF. In addition, the imaginary part of the IDF also furnishes very significant estimates of the
Raman (or electron) scattering cross-section. In what follows, we will analyze the IDF computed in two cases: keeping
the excitation energy (momentum transfer) fixed while varying the momentum transfer (excitation energy).

\begin{figure}[htbp]
\includegraphics*[width=8cm,height=9cm]{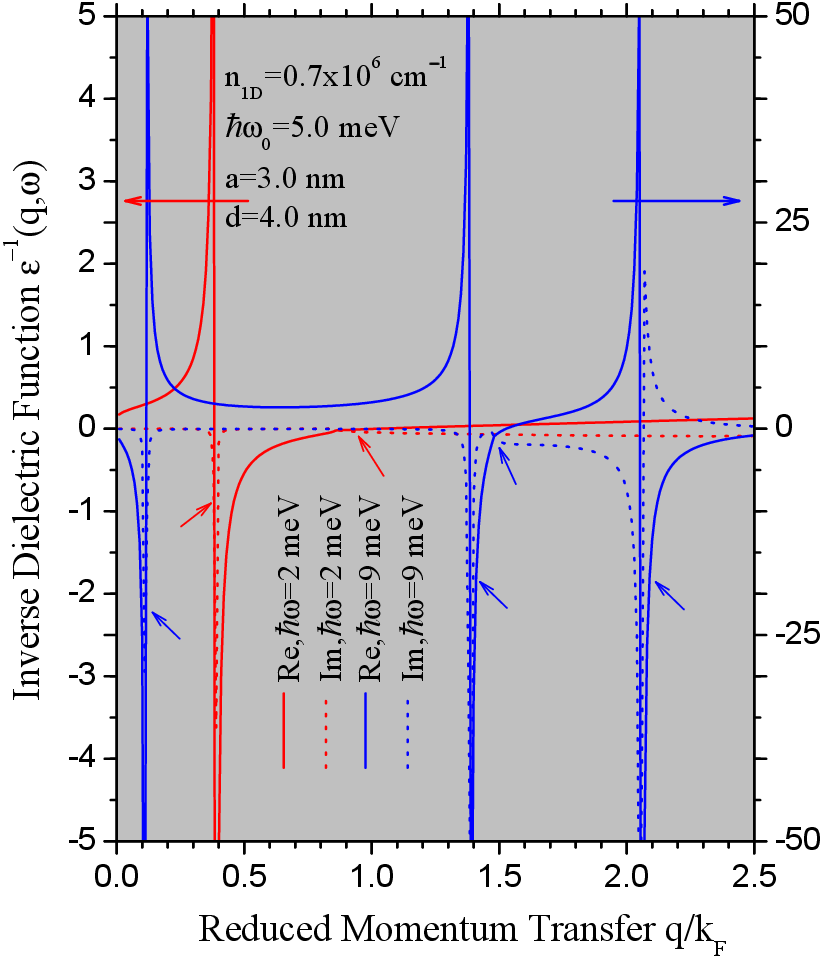}
\caption{(Color online) Inverse dielectric function $\epsilon^{-1}(q,\omega)$ vs. the momentum transfer $q/k_F$ for
the given values of the excitation energy $\hbar\omega$. The other parameters are as listed inside the picture.}
\label{fig16}
\end{figure}

Figure 16 illustrates the IDF $\epsilon^{-1}(q,\omega)$ as a function of the reduced momentum transfer $q/k_F$ for
two given values of the excitation energy: $\hbar \omega=2.0$ meV and 9.0 meV. It is worth mentioning that the quantity
that directly affects the transport phenomena is the spectral weight Im[$\epsilon^{-1}(q,\omega)$] which contains both
the single-particle contribution at large momentum transfer ($q$) and the collective (plasmon) contribution at small
$q$. Looking at Fig. 8 reveals that the resonances corresponding to the excitation energy $\hbar\omega=2.0$ (9.0) meV
should yield the intrasubband (intersubband) excitations. This is exactly what we observe. To be explicit, for $\hbar\omega=2.0$ meV, the sharp peak at $q/k_F=0.39$ produces the intrasubband collective (plasmon) mode while the
crossing of the real and imaginary parts at $q/k_F=0.87$ yields the intrasubband single-particle excitation. The
latter actually refers to the point on the upper edge of the intrasubband single-particle continuum. Similarly, for $\hbar\omega=9.0$ meV, the lowest, second lowest, and third lowest sharp peaks (counting from the origin) at
$q/k_F=0.11$, 1.39, and 2.06 reproduce, respectively, the intersubband collective (plasmon) mode before the dip, after
the dip, and the intrasubband plasmon within the intersubband single-particle continuum. The crossing of the real and
imaginary parts at $q/k_F=1.48$ yields exactly the point on the upper edge of the intersubband single-particle
continuum. Clearly, all these resonance peaks are a result of the existing poles of the IDF within the $\omega-q$ space.

\begin{figure}[htbp]
\includegraphics*[width=8cm,height=9cm]{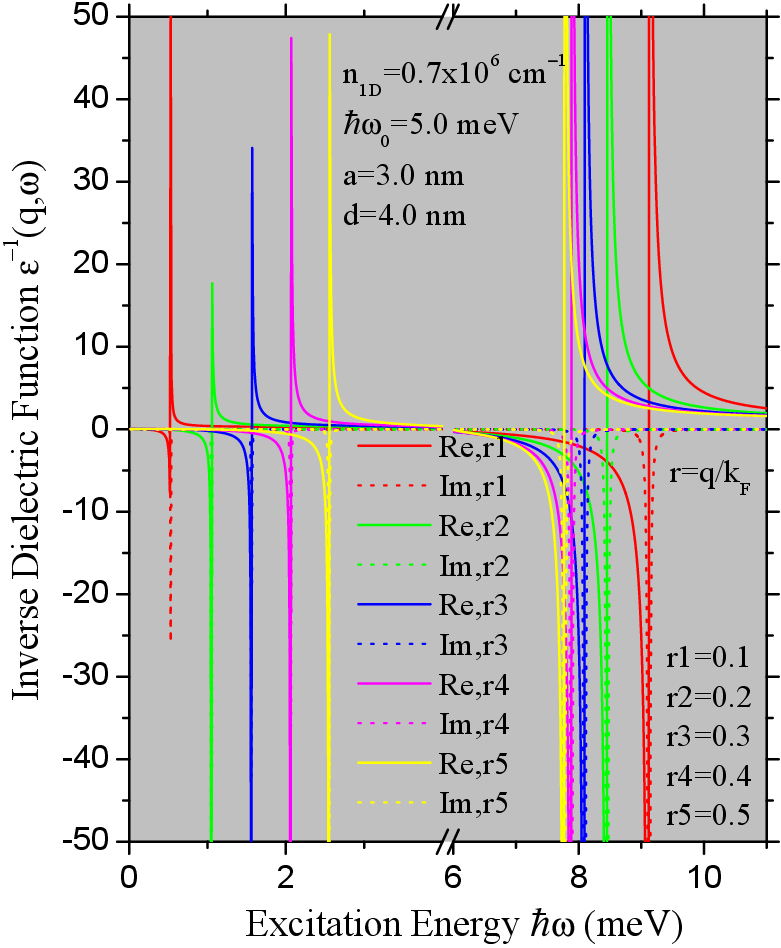}
\caption{(Color online) Inverse dielectric function $\epsilon^{-1}(q,\omega)$ vs. the excitation energy $\hbar\omega$
for the given values of the momentum transfer $q/k_F$. The other parameters are as listed inside the picture. Note the
scale break on the abscissa.}
\label{fig17}
\end{figure}

Figure 17 shows the IDF $\epsilon^{-1}(q,\omega)$ as a function of the excitation energy $\hbar\omega$ for the several
values of the momentum transfer: $q/k_F=0.1$, 0.2, 0.3, 0.4, and 0.5. Again, in analogy with the preceding discussion,
we focus on the imaginary part of the IDF. Since we choose relatively small values of the momentum transfer, we should
expect the resonance peaks to identify only the collective (plasmon) modes. We find two sharp peaks for a given
momentum transfer where the lower (upper) peak yields the intrasubband (intersubband) collective plasmon. To be specific,
the resonance peaks occurring at $\hbar\omega=0.53$, 1.06, 1.57, 2.07, and 2.56 reproduce the intrasubband plasmon and
those at  $\hbar\omega=9.13$, 8.46, 8.11, 7.89, and 7.77 yield the intersubband plasmon corresponding,
respectively, to the momentum transfer $q/k_F=0.1$, 0.2, 0.3, 0.4, and 0.5. Notice how the higher resonance peaks are
seen to exist in the {\em reverse} order to the values of $q$. This is simply because the intersubband plasmon, for
the chosen values of the $q$, propagates with the negative group velocity, until it observes the (roton-like) minimum.
For the moment, we do not want to expand on the rotonic character of the intersubband collective excitation observed
here and wish to leave it for the future.

\subsection{Comparison with the normal quantum wires}

A systematic investigation of the excitation spectrum -- comprised of single-particle and collective excitations --
in the quantum wires made up of vertically stacked quantum dots poses a question: how do we compare the coupled-dot
quantum wires (CDQW) with the normal quantum wires (NQW). In order to answer this question, we look at the excitation
spectra obtained in the two systems [see, e.g., Fig. 8 here and Fig. 133 in Ref. 1] and discuss some of the specific characteristics that distinguish the two systems apparently dubbed with the same nickname (i.e., quantum wires).
These are, for instance, the following. (a) The energy scales are usually higher in the NQW than in the CDQW. (b) The
intra- and inter-subband SPE (usually) overlap in the NQW unlike in the CDQW. (c) The minimum of the lower branch of
the intersubband SPE, generally, approaches (very close to) zero in the NQW unlike in the CDQW. (d) Both intra- and inter-subband CPE become Landau-damped at longer wavelengths in the NQW than in the CDQW. (e) The intersubband CPE, generally, starts to propagate with the positive group velocity in the NQW unlike in the CDQW. (f) The most important difference between the NQW and the CDQW is that the former offer us an (inherent) translational invariance (TI)
whereas the latter are imposed upon such TI (due, in fact, to the smaller length scales in the problem) for the sake
of (mathematical) convenience.

\section{Concluding Remarks}

In summary, we have investigated extensively the single-particle and collective (plasmon) excitations in the quantum
wires made up of vertically stacked (self-assembled) quantum dots within a two-subband model in the framework of the
Bohm-Pines' full RPA. The elementary characterization of such quantum wires by studying, e.g., the wave functions,
miniband structure, band width, density of states, and Fermi energy is followed by the principal results on the
single-particle and collective excitations, the influence of the well and barrier widths on the plasmon dispersion,
and the inverse dielectric functions. As to the intricacy of the methodology, we would like to stress that although
the TBA is, generally, labeled as the scheme accounting for the week tunneling effects, it is found to be reasonably
well successful. This is attributed to the smaller length scales involved in the problem.

As to the similarities and differences between the homogeneous quantum wires and the quantum wires made up of the
VSQD, we observe some important traits here. First of all, the presence of well and barrier widths provide us with
the freedom to tailor the excitation spectrum in desired energy range. Secondly, both intrasubband and intersubband
collective (plasmon) excitations are found to be free from Landau damping and comparatively long-lived in a greater
range of propagation vector. Thirdly, the intersubband plasmon excitation is observed to be originating and
propagating with a negative group velocity in the long wavelength limit. This is a roton character observed in the homogeneous quantum wires in the presence of an applied (perpendicular) magnetic field. We think this is an
important issue and requires a separate extensive study. It is interesting to see how inseparable is the role played
by the well and the barrier in influencing the plasmon propagation.

The motivation behind studying the inverse dielectric function is not solely to reaffirm the fact that the poles of
the IDF and the zeros of the DF yield exactly identical excitation spectrum, but also to pinpoint the advantage of
the former over the latter. For instance, the imaginary (real) part of the IDF sets to furnish a significant measure
of the longitudinal (Hall) resistance in the system. Moreover, the quantity Im [$\epsilon^{-1} (q,\omega)$] also
implicitly provides the reasonable estimates of the inelastic electron (or Raman) scattering cross-section $S$($q$)
for a given system.

Finally, we believe that the present investigation of the plasmon excitations in a quantum wire made up of VSQD is
of experimental importance given the excitement in the emerging fields of single-electron devices and the quantum computation. Considering the influence of an applied magnetic field, a confining potential that deviates from the
parabolic form, and the many-body effects could give better insight into the problem. We hope that such behavior characteristics of the plasmons as studied and predicted here can be verified by the Raman scattering experiments.




\begin{acknowledgments}
During the course of this work the author has benefited from many stimulating discussions and communications
with some colleagues. I would like to particularly thank B. Djafari-Rouhani, Naomi Halas, and Peter Nordlander.
Special thanks are due to Peter Nordlander for critical reading of the manuscript. I also wish to acknowledge
Kevin Singh for the generous help with the software throughout. It is a pleasure to thank Professor F. Barry
Dunning for all the support and encouragement.
\end{acknowledgments}

\newpage

\appendix

\section{The relevance of the Kronig-Penney model}

Here, we want to capture briefly the essence of the classic Kronig-Penney (KP) model with the Bastard's boundary
conditions (BBC) [44] incorporated. We do not want to pretend to be innovative since the KP model has so widely
been treated in the textbooks. Our intent is two-fold: the completeness of the eigenfunctions involved and to
uncover an interesting consequence of applying the Bastard's boundary conditions. The KP model is represented by
a 1D periodic potential shown in the right panel of Fig. 1. Even though the model is 1D, it is the periodicity of
the potential that is crucial to the resulting electronic band structure. The periodic potential in the
Schr\"{o}dinger equation is defined such as
\begin{equation}
V(z)=\left\{
\begin{array}{l}
V_0\, , \,\,\,\,\, {\rm if \,\,\,\,\, nd+a<z<(n+1)d} \\
0  \, , \,\,\,\,\,\,\, {\rm if \,\,\,\,\, nd<z<nd+a}
\end{array}
\right . \,
\end{equation}
As shown in Fig. 1, the potential has the period $d=a+b$. The Schr\"{o}dinger equation for this model is
\begin{eqnarray}
\frac{d^2 u}{dz^2} + \frac{2m^*_{w}\epsilon_0}{\hbar^2}\,u=0\, , \,\,\,\,\, {\rm if \,\,\,\,\, nd<z<nd+a}\\
\frac{d^2 u}{dz^2} - \frac{2m^*_{b}(V_0-\epsilon_0)}{\hbar^2}\,u=0\, , \,\,\,\,\, {\rm if \,\,\,\,\, nd+a<z<(n+1)d},
\end{eqnarray}
where $m^*_{w}$ ($m^*_{b}$) is the effective mass of the electron inside the well (barrier). The solutions to
these equations are
\begin{eqnarray}
u_1(z) = A e^{i\alpha z} + B e^{-i\alpha z}\, , \,\,\,\,\, {\rm if \,\,\,\,\, nd<z<nd+a}\\
u_2(z) = C e^{\beta (z-a)} + D e^{-\beta (z-a)}\, , \,\,\,\,\, {\rm if \,\,\,\,\, nd+a<z<(n+1)d},
\end{eqnarray}
where $\alpha=\sqrt{2m^*_{w}\epsilon_0/\hbar^2}$ and $\beta=\sqrt{2m^*_{b}(V_0-\epsilon_0)/\hbar^2}$ are real
quantities if we assume for the time being that $\epsilon_0<V_0$. Next, we need to match the boundary conditions
at $z=a$ and $z=d$ along with the Bloch theorem. These boundary conditions are:
\begin{eqnarray}
u_{1}(a) &=& u_{2}(a)\\
\frac{1}{m^*_w}\,u'_{1}(a) &=& \frac{1}{m^*_b}\,u'_{2}(a)\\
u_{1}(0) &=& e^{ikd}\,u_2(d)\\
\frac{1}{m^*_w}\,u'_{1}(0) &=& \frac{1}{m^*_b}\,e^{ikd}\,u'_{2}(d)
\end{eqnarray}
Eqs. (A7) and (A9) are becoming known as the BBC. A careful matching of these boundary conditions yields:
\begin{eqnarray}
A\,e^{i\alpha a}+B\,e^{-i\alpha a} &=& C + D\\
A\,e^{i\alpha a}-B\,e^{-i\alpha a} &=& S\,(C + D)\\
A+B &=& e^{ikd}\,[C\,e^{\beta b}+D\,e^{-\beta b}]\\
A-B &=& S\,e^{ikd}\,[C\,e^{\beta b}-D\,e^{-\beta b}],
\end{eqnarray}
where $k$ is the Bloch vector and $S=-i\,(\beta/m^*_b)/(\alpha/m^*_w)$. In order for there to be a nontrivial
solution to the Eqs. $(A10)-(A13)$ for $A$, $B$, $C$, and $D$, the determinant of their coefficients must vanish.
This yields the determinantal equation
\begin{equation}
\left |
\begin{array}{cccc}
e^{i\alpha a} \ & \ e^{-i\alpha a} \ & \ -1 \ & \ -1 \\
e^{i\alpha a} \ & \ -e^{-i\alpha a} \ & \ -S \ & \ S \\
1 \ & \ 1 \ & \ -e^{ikd}\,e^{\beta b} \ & \ -e^{ikd}\,e^{-\beta b} \\
1 \ & \ -1 \ & \ -S\,e^{ikd}\,e^{\beta b} \ & \ S\,e^{ikd}\,e^{-\beta b} \\
\end{array}
\right |=0
\end{equation}
It is straightforward, albeit lengthy, to simplify Eq. (A14) and prove that it is equivalent to
\begin{equation}
\cos(kd)=\cos(\alpha a)\,\cosh(\beta b) + i\,\frac{1+S^2}{2S}\,\sin(\alpha a)\,\sinh(\beta b)
\end{equation}
for $\beta$ real and
\begin{equation}
\cos(kd)=\cos(\alpha a)\,\cos(\gamma b) - \,\frac{1+S^2}{2S}\,\sin(\alpha a)\,\sin(\gamma b)
\end{equation}
for $\beta$ ($=i\gamma$) imaginary.
This is the classic (electron) dispersion relation for the 1D periodic system in the KP model, but with the
BBC embodied. For $m^*_w=m^*=m^*_b$, it exactly simplifies to the textbook result. Analytically, Eq. (A15)
may appear to be somewhat cumbersome, but it may be cast in a more transparent form. Let
$\alpha a=c_{1}\sqrt{\epsilon}$ and $\beta b=c_{2}\sqrt{1-\epsilon}$, where $\epsilon=\epsilon_0/V_0$,
$c_2=rc_1$,  $c_1=\sqrt{2m^{*}_{w} V_{0}/\hbar^2}\,a$, $r=r_l\sqrt{r_m}$, $r_l=b/a$, and $r_m=m^*_b/m^*_w$.
Then, Eq. (15) can be written in the form
\begin{equation}
\cos(kd)=
\cos(c_{1}\sqrt{\epsilon})\cosh(c_{2}\sqrt{1-\epsilon}) +
\frac{1-\epsilon (1+r_m)}{2\sqrt{r_m}\sqrt{\epsilon}\sqrt{1-\epsilon}}\,
\sin(c_{1}\sqrt{\epsilon})\sinh(c_{2}\sqrt{1-\epsilon})
\end{equation}
for $\epsilon < 1$ and
\begin{equation}
\cos(kd)=
\cos(c_{1}\sqrt{\epsilon})\cos(c_{2}\sqrt{\epsilon-1}) +
\frac{1-\epsilon (1+r_m)}{2\sqrt{r_m}\sqrt{\epsilon}\sqrt{\epsilon-1}}\,
\sin(c_{1}\sqrt{\epsilon})\sin(c_{2}\sqrt{\epsilon-1})
\end{equation}
for $\epsilon > 1$. In the limit of $\epsilon \rightarrow 1$, both Eqs. (16) and (17) yield, with $r_t=r_l\,r_m$,
\begin{equation}
\cos(kd)=\cos(c_1) - \frac{r_t c_1}{2}\sin(c_1),
\end{equation}
It is instructive to subject Eq. (15) to the limit of a homogeneous medium. This does not simply mean only to put $V_0=0$ but also to take $r_m=1$ in Eq. (15). In that case $\beta=i\alpha$ and Eq. (15) yields $\alpha=k$.
Therefore, we find $\epsilon_0=\hbar^2 k^2/2m^*$ for a free electron.


Next, we need to determine fully the wave functions in Eqs. $(A4)-(A5)$. For this purpose, we first write
the coefficients $B$, $C$, and $D$ in terms of $A$ from Eqs. $(A10)-(A13)$ and then apply the condition of
normalization
\begin{equation}
\frac{N}{2}\left\{\int^a_0 dz\,u^*_1(z)\,u_1(z) + \int^d_a dz\,u^*_2(z)\,u_2(z)\right\}=1,
\end{equation}
in order to determine $A$. Thus, the four coefficients $A$, $B$, $C$, and $D$ are defined as follows.
\begin{eqnarray}
A &=& \frac{1}{\sqrt{N}}
\left[a\,
\frac{\cos(\alpha a)\cos(kd)-\cosh(\beta b)}{\cos(\alpha a-kd)-\cosh(\beta b)} +
\frac{b}{t}\frac{\sin(\alpha a)}{\sinh(\beta b)}\,
\frac{\cos(\alpha a)-\cos(kd)\cosh(\beta b)}{\cos(\alpha a-kd)-\cosh(\beta b)}
\right]^{1/2}\\
B &=& -R^{-1}(1-S^2)\,\sinh(\beta b)\,A \\
C &=& r^{-1}R^{-1}(1+S)e^{-\beta b}[re^{-i\alpha a}-\cosh(\beta b)-\sinh(\beta b)]\,A\\
D &=& -r^{-1}R^{-1}(1-S)e^{\beta b}[re^{-i\alpha a}-\cosh(\beta b)+\sinh(\beta b)]\,A,
\end{eqnarray}
where $t=iS$, $r=e^{ikd}$, and $R=2S[re^{-i\alpha a}-\cosh(\beta b)]-(1+S^2)\sinh(\beta b)$. Thus the wave
functions $u_1(z)$ and $u_2(z)$ in Eqs. $(A4)-(A5)$ become finally known. The coefficient $A$ subjected to
the limit of a homogeneous medium and then substituting $k=0$ yields $A=1/\sqrt{Nd}$, just as expected. A
word of warning: you may never obtain the correct result if you reverse the order of the procedure, i.e. if
you substitute $k=0$ first and then apply the condition of homogeneity (i.e., $\beta=i\alpha$).

\newpage




\end{document}